\documentclass{aa}  

\usepackage{graphicx}
\usepackage{txfonts}
\usepackage{lipsum}
\usepackage{subcaption}         
\usepackage{lscape}             
\usepackage{placeins}           
                                
\usepackage[colorlinks=true,linkcolor=blue,citecolor=blue,urlcolor=blue]{hyperref}

\begin{document}

\nolinenumbers

\title{The Gaia-alerted 2024--2025 brightening of V346 Nor}

\author{\'A. K\'osp\'al\inst{1}
        \and
        D. J. Lengyel\inst{2}
        \and
        P. \'Abrah\'am\inst{1,3}
        \and
        B. M. Marosi\inst{1,4}
        \and
        F. Cruz-S\'aenz de Miera\inst{5}
        \and
        E. Fiorellino\inst{6}
        \and
        T. Giannini\inst{7}
        \and
        S. Kraus\inst{8}
        \and
        A. Labdon\inst{9}
        \and
        P. W. Lucas\inst{10}
        \and
        Th. Moore\inst{11}
        \and
        Zs. Nagy\inst{1}
        \and
        M. Siwak\inst{12,1}
        \and
        A. Udalski\inst{13}
       }

   \institute{Konkoly Observatory, HUN-REN Research Centre for Astronomy and Earth Sciences, MTA Centre of Excellence, Konkoly-Thege Mikl\'os \'ut 15-17, 1121 Budapest, Hungary
        \and
        School of Physics and Astronomy, University of Glasgow, Glasgow G12 8QQ, UK
        \and
        Institute for Astronomy (IfA), University of Vienna, T\"urkenschanzstrasse 17, A-1180 Vienna, Austria
        \and
        Institute of Physics and Astronomy, ELTE E\"otv\"os Lor\'and University, P\'azm\'any P\'eter s\'et\'any 1/A, 1117 Budapest, Hungary
        \and
        Universit\'e de Toulouse, CNES, CNRS, IRAP, 9 avenue du Colonel Roche, 31028 Toulouse Cedex 4, France
        \and
        Alma Mater Studiorum---Universit\`a di Bologna, Dipartimento di Fisica e Astronomia ``Augusto Righi,'' Via Gobetti 93/2, I-40129 Bologna, Italy
        \and
        INAF---Osservatorio Astronomico di Roma, via di Frascati 33, 00078, Monte Porzio Catone, Italy
        \and
        School of Physics and Astronomy, University of Exeter, Exeter EX4 4QL, UK
        \and
        European Southern Observatory, Alonso de Cordoba 3107, Vitacura, Santiago, Chile
        \and
        Centre for Astrophysics Research, University of Hertfordshire, College Lane, Hatfield, AL1 09A, UK
        \and
        Astrophysics Research Centre, School of Physics and Astronomy, Queen's University, Belfast, BT7 1NN, UK
        \and
        Mt. Suhora Astronomical Observatory, University of the National Education Commission, ul. Podchor\k{a}\.zych 2, 30-084 Krak{\'o}w, Poland
        \and
        Astronomical Observatory, University of Warsaw, Al. Ujazdowskie 4, 00-478 Warszawa, Poland
             }

   \date{Received date; accepted date}

  \abstract
   {Young stellar objects (YSOs) build their final mass through a highly unsteady, episodic accretion process, often accompanied by outflow and ejection events. Objects undergoing episodic high-accretion states are known as eruptive YSOs. }
   {V346\,Nor is an embedded eruptive YSO that underwent a sudden optical brightening in 2024 July that triggered an alert from the Gaia Photometric Science Alert system. With the aim of characterizing the physical processes behind this recent brightening event, we analyze new and archival photometric and spectroscopic monitoring data.} 
   {We study how the brightness, colors, and spectral features changed with time in the past decade. We also fitted an accretion disk model to the observations in which the stellar mass--accretion rate product $M_{\star}\dot{M}$ and the line-of-sight extinction $A_V$ are free parameters.}
    {The accretion rate in V346\,Nor has been higher ($M_{\star}\dot{M} \sim 10^{-4}\,M_{\odot}^2\,$yr$^{-1}$) since 2021 than during any previous historical peak, although it was relatively constant during the 2025 February-May brightness maximum, which was driven instead by a drop in extinction from $A_V \approx 20$\,mag to $10$\,mag. V346\,Nor has accreted about 2\% of the stellar mass since its outburst. A P\,Cygni profile in H$\alpha$ indicates a strong outflowing wind reminiscent of classical FUors. The variability of the CaII infrared triplet lines and transient CO bandhead emission point to a highly unstable inner disk and accretion region. The persistence of forbidden line emission indicates ongoing jet activity.}
   {The brightness and color variations of V346 Nor are the result of a superposition of changes in the accretion rate and the line-of-sight extinction. With record-high accretion levels and strong wind and jet activity, V346\,Nor is in one of its most active phases since its original outburst in the 1970s.}

   \keywords{Stars: pre-main sequence --
                stars: individual: V346 Nor --
                protoplanetary disks --
                accretion, accretion disks -- 
                infrared: stars
               }

\maketitle
\nolinenumbers


\section{Introduction}

Young stellar objects (YSOs) build their final mass in an unsteady process that involves episodic (eruptive) accretion from the circumstellar disk onto the central object, often accompanied by repeated outflow and ejection events as well \citep{reipurth2001, fischer2023}. The physical driver of these eruptions is thought to be thermal or gravitational instabilities in the circumstellar disk, which cause a runaway increase in the accretion rate that can persist for years to decades \citep[e.g.,][]{bell1994, vorobyov2006}. Eruptive YSOs are broadly classified by their outburst amplitudes, timescales, and spectroscopic properties.

FU\,Orionis objects (FUors)  brighten by 4–6 magnitudes over months to years and remain elevated for decades, with accretion rates of $10^{-5} - 10^{-4},M_{\odot}\,yr^{-1}$ \citep[e.g.,][]{szabo2021}, probably proceeding through a boundary layer \citep{popham1996, labdon2021, carvalho2024}. EX\,Lupi-type objects (EXors) undergo shorter, repetitive bursts of 1–3 magnitudes over weeks to months, but their spectra are consistent with magnetospheric accretion operating at an elevated rate \citep{sicilia2012, sicilia2015, pouilly2024}. V1647 Ori-type objects (MNors) bridge these two classes in terms of timescales, variability amplitudes, and spectral properties. The same magnetospheric accretion mechanism works in them as in EXors, but they likely represent a younger, more deeply embedded evolutionary stage \citep{hamaguchi2012, contreraspena2017}. An up-to-date list of currently known young eruptive stars is available in the OYCAT\footnote{\href{http://starformation.synology.me:5002/OYCAT/main.html}{http://starformation.synology.me:5002/OYCAT/main.html}} catalog \citep{contreraspena2025}.

YSOs currently accreting at extremely high rates may be {\it bona fide} FUors (if an outburst was observed), FUor-like objects (if the brightening was not directly observed), or peculiar objects (if they show only some of the properties typical of FUors). Their optical and infrared spectra show a characteristic multi-temperature structure, dominated by absorption features forming in the cooler atmosphere of a hot, optically thick accretion disk, including CO bandhead absorption in the near-infrared and P Cygni profiles in H$\alpha$ \citep{connelley2018, portnoi2026}.

V346\,Nor (located in the Sa\,187 molecular cloud at a distance of 700\,pc, \citealt{alvarez1986}) is a YSO that went into outburst in the 1970s and have peculiar spectral characteristics probably thanks to its embedded nature and strong jet \citep{reipurth1985,reipurth1997}. It reached a broad maximum brightness between 1990 and 2000 and displayed a deep minimum between 2008 and 2013. This fading was interpreted by \citet{kraus2016} and \citet{kospal2017a} as a temporary halt in the accretion rate. The peak accretion rate of V346\,Nor was $10^{-4}\,M_{\odot}$\,yr$^{-1}$, reached in 1992, while during the minimum in 2010, this value went below $4\times10^{-7}\,M_{\odot}$\,yr$^{-1}$. Using ALMA data from 2014-2015, \citet{kospal2017c} measured the radial velocity profile in the disk and pseudo-disk and found that the infall rate in the pseudo-disk is $6\times10^{-6}\,M_{\odot}$\,yr$^{-1}$. They concluded that the mismatch between the disk-to-star accretion and the envelope-to-disk infall may be the cause of the eruptive nature of the object.

The optical and near-infrared spectra of FUors is dominated by the inner active accretion disk \citep{liu2022}. FUors typically have absorption features owing to the temperature profile in the active accretion disk, which is hot in the midplane and colder on the surface, the opposite of what is usual for regular, normally accreting T Tauri stars. V346\,Nor is peculiar in this respect, because it displays emission lines as well. Most emission lines in its spectrum are forbidden metallic and H$_2$ lines, typical jet tracers \citep{kospal2020}. Such lines are observable not only toward the source, but also about $5''$ away to the south, at the position of the Herbig-Haro object HH\,57 \citep{reipurth1997}, indicating both more recent and older ejection activities. 

V346\,Nor underwent a sudden optical brightening in 2024 July that triggered an alert from the Gaia Photometric Science Alert system of the Gaia astrometric space mission \citep{hodgkin2021}, where it appeared under the name Gaia24chn. Motivated by this alert, here we present results based on new and archival photometric and spectroscopic monitoring data from various optical and near-infrared instruments and draw conclusions about the physical processes behind this recent brightening. In Sec.~\ref{sec:obs} we present details of the observations we obtained for V346\,Nor and their reduction. In Sec.~\ref{sec:res} we present the resulting light curves, spectral energy distributions (SEDs), analyze the variability spectral features and fit an accretion disk model to the data. In Sec.~\ref{sec:sum}, we summarize our conclusions. 


\section{Observations and data reduction}
\label{sec:obs}

\subsection{Optical photometry}

We obtained optical photometric monitoring of V346\,Nor using the 60\,cm diameter Rapid Eye Mount (REM) telescope, located at La Silla Observatory (Chile) between 2021-Apr-06 and 2024-Sep-22. In 2021, we used a strict monthly cadence, while in 2023 and 2024, data were taken at every few days to few weeks. The optical camera on REM, ROS2 is equipped with Sloan $griz$ filters and has a field of view of $9\farcm1\times9\farcm1$ and a pixel scale of $0\farcs58$. We typically obtained 5--15 frames per filter using an exposure time of 60\,s in all filters. After the usual bias subtraction and flat fielding, we shifted and coadded the images obtained with the same filter on the same night and calculated aperture photometry for V346\,Nor and 12 comparison stars in the field of view with various apertures with radii between 1 and 5 pixels and sky annulus between 10 and 20 pixels. Photometric calibration was done using the APASS 9 magnitudes of the comparison stars without fitting any color term. We finally averaged the calibrated magnitudes obtained with the different aperture radii. V346\,Nor remained undetected in $g$ band throughout our monitoring while in $r$ until 2024 April; in these cases we give 1$\sigma$ upper limits.

We complemented our REM light curves with data downloaded from the ATLAS forced photometry server\footnote{\href{https://fallingstar-data.com/forcedphot/}{https://fallingstar-data.com/forcedphot/}}. ATLAS (Asteroid Terrestrial-impact Last Alert System) is an asteroid impact early warning system that makes images of the full sky several times per night using two telescopes in Hawai'i, one in Chile, and one in South Africa \citep{tonry2018}. The telescopes have 0.65\,m mirror diameter with 0.5\,m Schmidt correctors, and use two broad-band filters: orange ($o$, covering the $560-820\,$nm range) and cyan ($c$, covering the $420-650\,$nm range). ATLAS covers the whole observable sky every two days, obtaining 4 images per night of a specific region/object.

V346\,Nor was observed in the framework of the Optical Gravitational Lensing Experiment (OGLE; \citealt{udalski2015}) under the name gd1112.05.9732. OGLE data for V346\,Nor obtained until 2019-Sep-8 is published in \citet{kospal2020}. Here we present data obtained since then until 2025-Jun-16. These observations were taken in the $I_C$ band using the 1.3\,m Warsaw Telescope at Las Campanas Observatory in Chile with an approximately 1-3\,days cadence. Details of the OGLE instrumental setup, data reduction, and calibration can be found in \citet{udalski2015}.

The Gaia $G$ band light curve of V346\,Nor is available from the Gaia alerts website\footnote{\href{http://gsaweb.ast.cam.ac.uk/alerts/alert/Gaia24chn/}{http://gsaweb.ast.cam.ac.uk/alerts/alert/Gaia24chn/}} and contains 21 valid Gaia $G$-band magnitudes taken between 2016 August and 2024 August.

\begin{figure*}
\centering
\includegraphics[width=\textwidth]{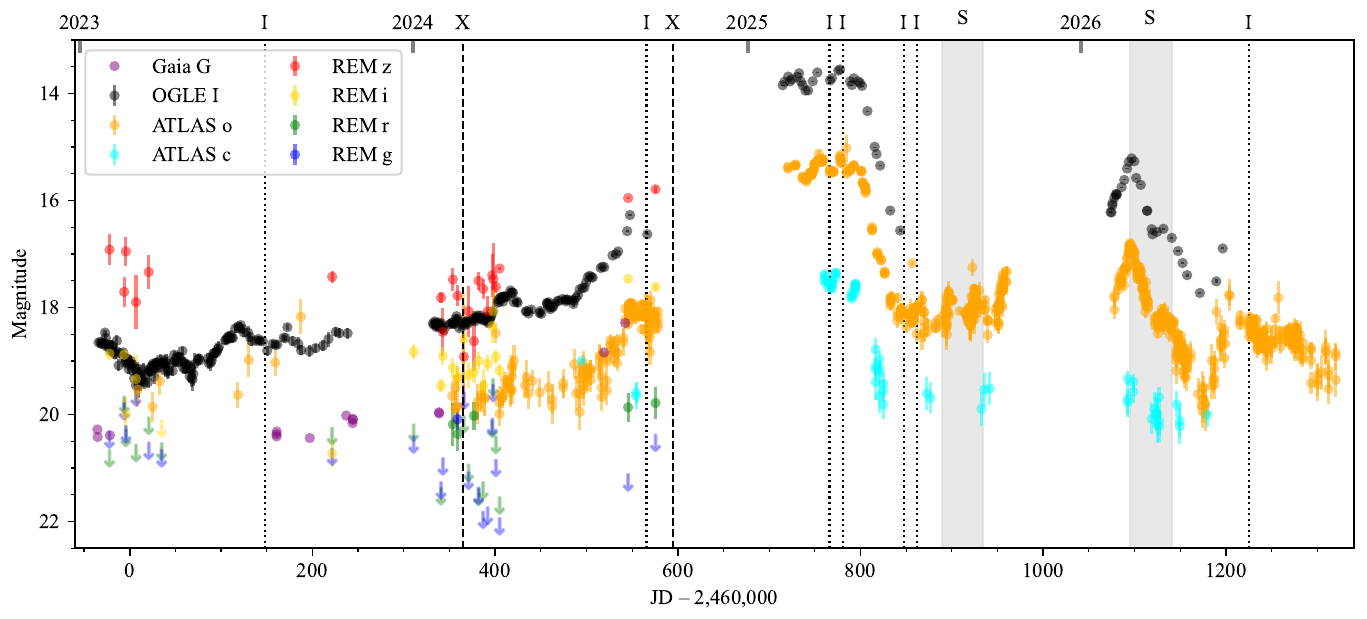}
\caption{Optical light curves of V346\,Nor in the 2023--2026 time period. Vertical dotted (dashed) lines mark the epochs of IRTF/SpeX (VLT/XSHOOTER) spectroscopy, while the shaded gray regions mark when SPHEREx data were taken.}\label{fig:light}
\end{figure*}

The results of our optical REM photometry are listed in Tab.~\ref{tab:phot} and all optical photometric data from the 2023--2026 period are plotted together in Fig.~\ref{fig:light}.

\subsection{Infrared photometry}

We monitored V346\,Nor between 2021-Apr-6 and 2024-Sep-22 in the near-infrared, using REMIR, the infrared camera on the REM telescope. REMIR is equipped with $JHK'$ filters and has a field of view of $10'\times10'$ and a pixel scale of $1\farcs2$. The exposure times varied considerably ($J$: $5-60$\,s; $H$: $0.8-30$\,s; $K_S$: $5$\,s), as part of the continuous optimization of the observing strategy to the extreme brightness variations of the source over the observing period. The images were reduced with our own IDL routines, including the construction and subtraction of a sky image, flat-fielding, and the production of a mosaic from the 5 dither positions. Then we extracted instrumental magnitudes for the target as well as for 4-14 bright 2MASS stars in the mosaic frame. We selected these comparison stars so that they have the highest 2MASS quality flags of A \citep{cutri2003}. We performed aperture photometry with an aperture radius of 3 pixels ($3\farcs7$). Finally, on each night we determined an average offset between the instrumental and the 2MASS magnitudes of the comparison stars, and used it to calibrate V346\,Nor. No color term was included in the calibration. 

The position of V346\,Nor (RA$_{\rm J2000}$=16$^{\rm h}$32$^{\rm m}$32.19$^{\rm s}$, DE$_{\rm J2000}$=--44$^{\circ}$ 55$'$ 30$\farcs$66) was covered by the VISTA Variables in the V\'\i{}a L\'actea extended (VVVX) ESO public survey \citep{saito2024}, with the $K_S$ filter. We obtained photometry on 2021-Feb-14 and 2022-Aug-01, following the reduction and calibration steps, including a linearity correction, described in Appendix~A of \citet{kospal2017a}. 

As a supplement to the near-infrared spectroscopy carried out with the IRTF (Sect.~\ref{sec:spec}), we obtained $JHKL'$ band imaging using the internal guide camera of the SpeX instrument. It is a 512$\times$512 Aladdin 2 InSb array, with a pixel size of $0\farcs12$, and a field of view of approximately $1'\times1'$. We adopted a 5-point dither pattern, with typical exposure times of 40\,s in $J$, 5\,s in $H$, 2\,s in $K$, and 0.5s in $L'$. For comparison, we observed a nearby bright star (HD328329, HD148865, or J16330061-4457300). Following the construction and subtraction of a sky image, we built a mosaic from each dither sequence. No flatfielding was applied to the frames. We performed aperture photometry using a large aperture radius of 40 pixel, which included the full point-spread function (PSF), and was thus insensitive for temporal variations in the PSF shape. The final calibrated photometry was computed by adding the standard 2MASS ($JHK_S$ bands) or AllWISE $W1$ ($L'$-band) magnitude to the difference of the instrumental magnitudes of V346\,Nor and the comparison star. Note that while for the observations we used the guide camera filters matching the Mauna Kea Observatories Near-Infrared Photometric System (MKO), the standard magnitudes were taken from the 2MASS or the WISE system. Thus, we consider our final results also in these systems, ignoring the possible differences in filter profiles. 

The results of our near-infrared REM, VISTA, and IRTF/guidedog photometry are listed in Tab.~\ref{tab:phot}.

The Wide-field Infrared Survey Explorer (WISE) also measured V346\,Nor both during its cryogenic mission and during the NEOWISE reactivation \citep{wright2010, mainzer2014} using the $W$1 and $W$2 (3.35\,$\mu$m and 4.60\,$\mu$m, respectively) filters. V346\,Nor strongly saturated WISE, therefore we applied saturation bias correction as described in section II.1.c.iv.a. of the Explanatory Supplement to the NEOWISE Data Release Products\footnote{\href{https://wise2.ipac.caltech.edu/docs/release/neowise/expsup/sec2_1civa.html}{https://wise2.ipac.caltech.edu/docs/release/neowise/expsup/\\sec2\_1civa.html}}. Finally, we calculated seasonal (half-year) averages, as the WISE satellite scanned the whole sky every half year. The results are listed in Tab.~\ref{tab:wise}

\begin{figure*}
\centering
\includegraphics[width=\textwidth]{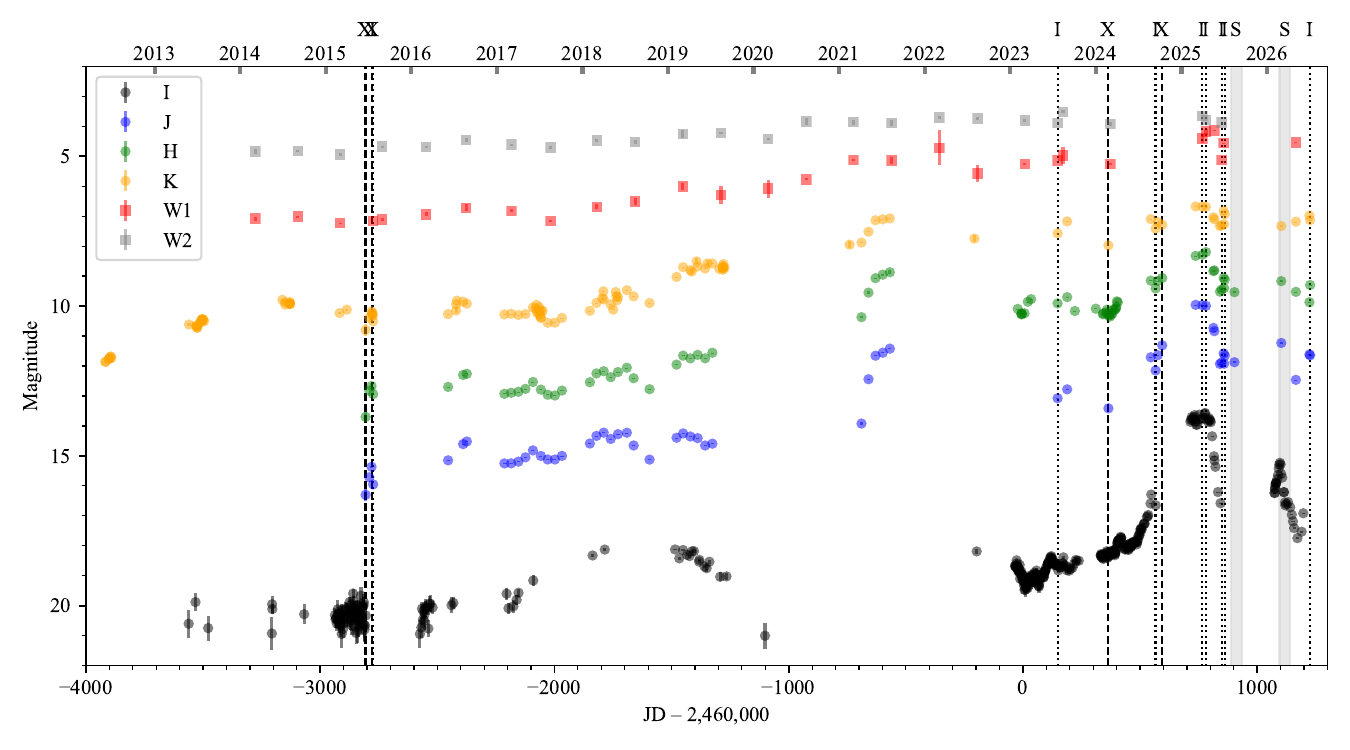}
\caption{Near-infrared and mid-infrared light curves of V346\,Nor in the 2012-2026 time period (along with the OGLE $I_C$ data for comparison). Vertical dotted (dashed) lines mark the epochs of IRTF/SpeX (VLT/XSHOOTER) spectroscopy, while the shaded gray regions mark when SPHEREx data were taken.}\label{fig:light_nir}
\end{figure*}

All near-infrared ($JHK_S$) and mid-infrared ($W$1 and $W$2) photometric data from 2012 -- 2026 period are plotted in Fig.~\ref{fig:light_nir}, together with the OGLE $I_C$ light curve for reference.

\subsection{Spectroscopy}
\label{sec:spec}

We obtained optical-infrared medium resolution spectra of V346\,Nor using the SpeX spectrograph \citep{rayner2003} at the 3.2\,m diameter NASA Infrared Telescope Facility (IRTF) located on Mauna Kea, Hawai'i (USA) on 2023-Jul-22, 2024-Sep-12, 2025-Mar-31, 2025-Apr-15, 2025-Jun-21, 2025-Jul-5, and 2026-Jul-4. We also downloaded an earlier spectrum of V346\,Nor from the IRTF archive, obtained on 2015-Jul-22, previously published in \citet{connelley2018}. Tab.~\ref{tab:log} shows our spectroscopic log of observations, including information about the grating, slit width, and telluric standard used in each observation. The SXD grating covers the $0.7-2.55\,\mu$m range and provides a spectral resolution of $R=1200$ with the $0\farcs5$ slit and 750 with the $0\farcs8$ slit. The LXD\_short covers $1.67-4.2\,\mu$m, while the LXD\_long covers $1.98-5.3\,\mu$m and they both provide a resolution of $R=1500$ with the $0\farcs5$ slit.

\begin{figure*}
\centering
\includegraphics[width=\textwidth]{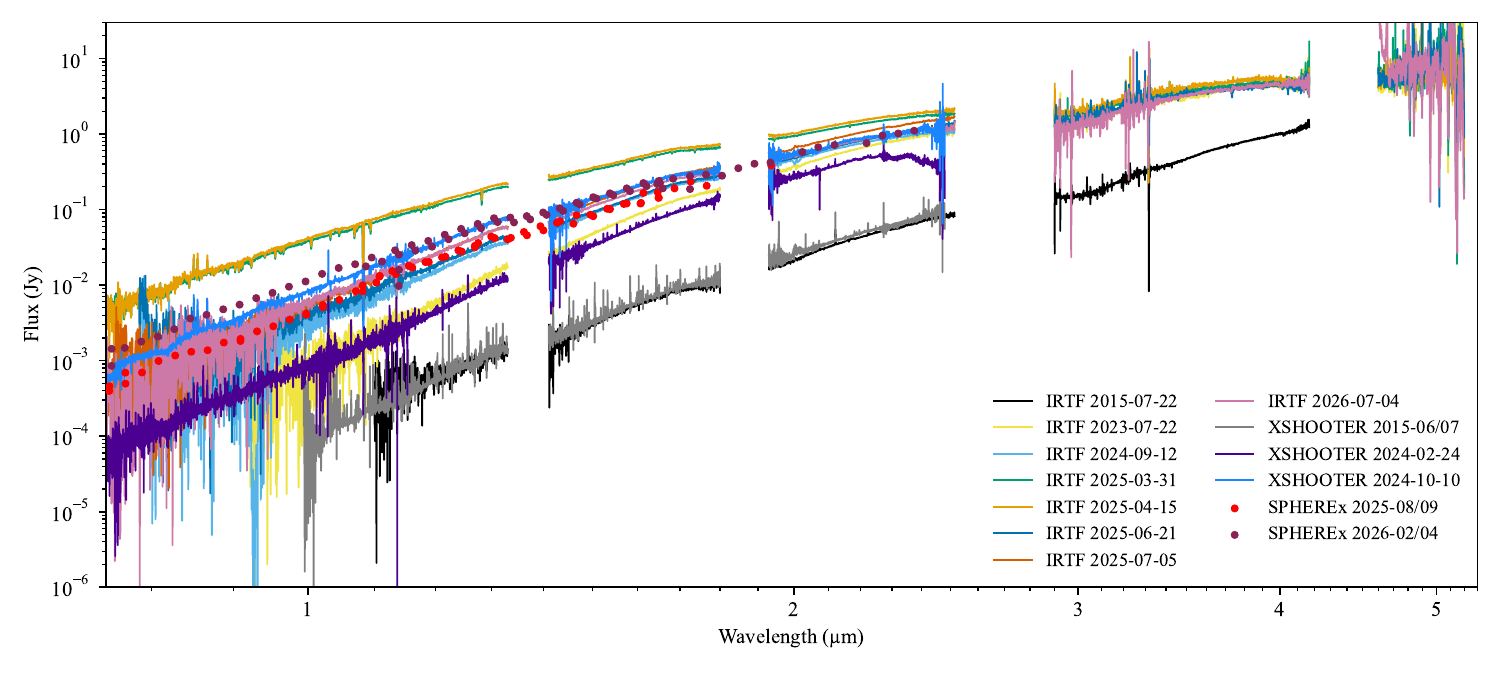}
\caption{Optical-infrared spectra of V346\,Nor obtained between 2015 and 2026 with the VLT/XSHOOTER and IRTF/SpeX spectrographs, as well as with the SPHEREx mission (after discarding the saturated data points).}\label{fig:spec}
\end{figure*}

We reduced all SpeX spectra in the same way using SpexTool \citep{vacca2003,cushing2004}. Data reduction steps included flat correction, subtraction of A and B nod positions, calculation of the spectroscopic traces in the 2D images, extraction of the 1D spectra in tapered apertures, division by the telluric spectra to correct for instrumental response and telluric absorption and emission, and the stitching of the various \'echelle orders. The telluric spectra were also used for absolute flux calibration. The reduced, flux calibrated spectra are displayed in Fig.~\ref{fig:spec}. We also derived synthetic photometry by multiplying the flux-calibrated spectra with the respective filter profiles, summing up the obtained values, and converting them from Jy to Vega magnitude. Finally, we scaled the spectra so that synthetic $JHK_S$ photometry, obtained by multiplying the spectra with the 2MASS $JHK_S$ filter profiles and integrating, matched the guidedog photometry taken on the same night. This scaling, which provides more accurate flux calibration than the telluric standard alone, was done for all epochs except for 2024-Sep-12, when no guidedog images were taken of V346\,Nor.

We obtained optical-infrared medium resolution spectra of V346\,Nor using XSHOOTER \citep{vernet2011} on the 8.2\,m diameter UT3 telescope of the Very Large Telescope (VLT) located in Paranal Observatory (Chile) on 2024-Feb-24 and 2024-Oct-10. We used the $0\farcs8$, $0\farcs7$, and $0\farcs6$ wide slits in the UVB, VIS, and NIR arms covering the $300-559.5$\,nm, $559.5-1024$\,nm, and $1024-2480$\,nm wavelength ranges with spectral resolution of 6700, 11400, and 8100, respectively. We reduced the observations using the ESO X-Shooter pipeline V.3.5.3 within the EsoReflex environment (CPL-7.3.2). We corrected for telluric absorption using the Molecfit software \citep{kausch2015,smette2015}. In 2024 October we also obtained spectra with the $5''$ wide slit that provided spectra without slit losses. We scaled the narrow slit spectra to the flux level of the wide slit measurements to obtain a flux-calibrated spectrum. In 2024 February only narrow slit spectra were available. In order to correct for potential flux losses, we computed synthetic photometry from these spectra in the optical I$_C$ and near-infrared $H$ bands, and determined scaling factors comparing these synthetic magnitudes with our OGLE I$_C$ (Sect.~2.1) and REMIR $H$-band (Sect.~2.2) measurements, interpolating for the date of the X-Shooter observations. Tab.~\ref{tab:log} shows the total exposure times used during the observations.

V346\,Nor was also observed with the SPHEREx (Spectro-Photometer for the History of the Universe, Epoch of Reionization and Ices Explorer) observatory \citep{bock2026}. SPHEREx is a NASA  Medium-Class Explorers mission designed to perform an all-sky spectral survey. SPHEREx has a wide-field telescope whose effective aperture is 20\,cm, and for spectroscopy it uses a Linear Variable Filter (LVF) which collects both spectral and imaging data simultaneously in over 100 near-infrared bands. Reduced spectra can be accessed through the SPHEREx Spectrophotometry Tool at  the \href{https://irsa.ipac.caltech.edu/Missions/spherex.html}{NASA/IPAC Infrared Science Archives} (IRSA). The spectra cover the 0.75 -- 5.0 $\mu$m wavelength range at low resolution, with $R$ = 35 to 128. V346\,Nor was observed with SPHEREx twice, once between 2025 August and September and once again between 2026 February and April. Data points above $\sim1.5\,\mu$m are over the saturation limit according to the pre-launch performance model (see Table 17 in the SPHEREx Explanatory Supplement Version 1.3\footnote{\href{https://irsa.ipac.caltech.edu/data/SPHEREx/docs/SPHEREx_Expsupp_QR.pdf}{https://irsa.ipac.caltech.edu/data/SPHEREx/docs/\\SPHEREx\_Expsupp\_QR.pdf}}), although data points seem ``well-behaved'' until about 2$\,\mu$m. A higher scatter of data points, indicative of more severe saturation effects, are visible above $\sim2.4\,\mu$m. This prompted us to check the saturation limits of the mission based on actual data and not rely blindly on the pre-launch model. Our results, summarized in Appendix~\ref{sec:spherex}, indicate that data about 0.5-1\,mag brighter than the  pre-launch saturation levels are still as reliable as data just below these flux levels. For V346\,Nor, we will use only data below $\sim2.4\,\mu$m in wavelength and below $\sim1\,$Jy in flux.

\begin{table*}
\caption{Log of Spectroscopic Observations of V346\,Nor}
\label{tab:log}
\centering
\begin{tabular}{c c c c c c c}
\hline\hline
Date & JD & Telescope/Instrument & Slit & Grating / arm & Exp.~time & Telluric standard \\
\hline
2015-Jul-22 & 2457225.75 & IRTF/SpeX    & $0\farcs5\times15''$ & SXD        & 2157\,s & HD 154410 \\
2015-Jul-22 & 2457225.75 & IRTF/SpeX    & $0\farcs5\times15''$ & LXD\_short & 467\,s  & HD 154410 \\
2023-Jul-22 & 2460147.79 & IRTF/SpeX    & $0\farcs5\times15''$ & SXD        & 598\,s  & HD 154410 \\
2023-Jul-22 & 2460147.79 & IRTF/SpeX    & $0\farcs5\times15''$ & LXD\_long  & 278\,s  & HD 154410 \\
2024-Feb-24 & 2460364.86 & VLT/XSHOOTER & $0\farcs8\times11''$ & UVB        & 1800\,s & \dots  \\
2024-Feb-24 & 2460364.86 & VLT/XSHOOTER & $0\farcs7\times11''$ & VIS        & 1800\,s & \dots \\
2024-Feb-24 & 2460364.86 & VLT/XSHOOTER & $0\farcs6\times11''$ & NIR        & 500\,s  & \dots  \\
2024-Sep-12 & 2460565.74 & IRTF/SpeX    & $0\farcs8\times15''$ & SXD        & 297\,s  & HD 154196 \\
2024-Oct-10 & 2460594.49 & VLT/XSHOOTER & $0\farcs8\times11''$ & UVB        & 3000\,s & \dots \\
2024-Oct-10 & 2460594.49 & VLT/XSHOOTER & $0\farcs7\times11''$ & VIS        & 2900\,s & \dots \\
2024-Oct-10 & 2460594.49 & VLT/XSHOOTER & $0\farcs6\times11''$ & NIR        & 1200\,s & \dots \\
2025-Mar-31 & 2460766.10 & IRTF/SpeX    & $0\farcs5\times15''$ & SXD        & 1494\,s & HD 154410 \\
2025-Mar-31 & 2460766.13 & IRTF/SpeX    & $0\farcs5\times15''$ & LXD\_long  & 556\,s  & HD 154410 \\
2025-Apr-15 & 2460781.04 & IRTF/SpeX    & $0\farcs5\times15''$ & SXD        & 717\,s  & HD 154410 \\
2025-Apr-15 & 2460781.05 & IRTF/SpeX    & $0\farcs5\times15''$ & LXD\_long  & 510\,s  & HD 154410 \\
2025-Jun-21 & 2460847.85 & IRTF/SpeX    & $0\farcs5\times15''$ & SXD        & 474\,s  & HD 154410 \\
2025-Jun-21 & 2460847.84 & IRTF/SpeX    & $0\farcs5\times15''$ & LXD\_long  & 334\,s  & HD 154410 \\
2025-Jul-05 & 2460861.82 & IRTF/SpeX    & $0\farcs8\times15''$ & SXD        & 2093\,s & HD 144399 \\
2025-Aug-15$^*$ & 2460902.99$^*$ & SPHEREx & \dots & \dots & 112.5\,s & \dots \\
2026-Jul-04 & 2461225.81 & IRTF/SpeX    & $0\farcs8\times15''$ & SXD        & 474\,s & HD 154410 \\
2026-Jul-04 & 2461225.82 & IRTF/SpeX    & $0\farcs8\times15''$ & LXD\_long  & 334\,s & HD 154410 \\
2026-Mar-03$^*$ & 2461102.73$^*$ & SPHEREx & \dots & \dots & 112.5\,s & \dots \\
\hline
\end{tabular}
\tablefoot{$^*$Average date and JD. SPHEREx observations of V346\,Nor were made between 2025-Aug-02 -- 2025-Sep-15 and again between 2026-Feb-23 -- 2026-Apr-10.}
\end{table*}

\section{Results and analysis}
\label{sec:res}

\subsection{Photometric brightness and color evolution}

Figure~\ref{fig:light} shows the optical light curves of V346\,Nor in the 2023--2026 time period. Based mostly on the OGLE $I_C$ and Atlas $o$ curves, V346\,Nor displayed a slow, gradual brightening throughout 2023 and the first half of 2024, followed by a steep rise starting in mid-2024. This triggered the Gaia alert mentioned before, on 2024-Jul-27. By the time it was past conjunction with the Sun, in 2025 February, V346\,Nor reached a plateau with smaller ($<$0.4\,mag) fluctuations. This was followed by a steep decline between 2025 May-June to another, lower level, but still above the brightness before mid-2024.

Atlas $o$ data from the new observing season in 2026 revealed that V346\,Nor again displays significant brightness variations, with peak-to-peak amplitude of about $\Delta o = 3.2$\,mag. In this period, the peak brightness occurred around 2026-Feb-23/25 at $o=16.74\,$mag, while the minimum was on 2026-May-17 at around $o=19.90\,$mag. This minimum magnitude is close to how faint V346\,Nor was in the first half of 2024, just before the Gaia-alerted event.

Figure~\ref{fig:light_nir} shows the near-infrared ($JHK_S$) and mid-infrared ($W$1 and $W$2) light curves of V346\,Nor, showing the brightness evolution after its deep minimum in 2010-2011 \citep{kospal2020}. While there is a general similarity in the light curves at various wavelengths, further scrutiny reveals a complicated behavior. The $H$ band and longer wavelength data show a mostly constant plateau in 2015-2018, while the $I$ band curve shows that a brightening started in the optical in early 2017. In 2019, the $H$ and $K_S$ curves show brightening while the $I$ and $J$ curves show fading. In 2021, a sharp, blue rise happened, with significantly higher amplitudes in the $J$ than in the $K_S$ band.

V346\,Nor was rather stable in brightness from early 2023 to early 2024: the $H$-band data display only smaller ($< 1\,$mag) fluctuations during this period. The brightness peak that triggered the Gaia alert is well visible in the near-infrared light curves as well. V346\,Nor brightened in $JHK_S$ in 2024, had a plateau in early 2025, followed by a fading. The variability amplitudes, however, are smaller than in the optical, and decrease with increasing wavelength.

\begin{figure*}
\centering
\includegraphics[width=\textwidth]{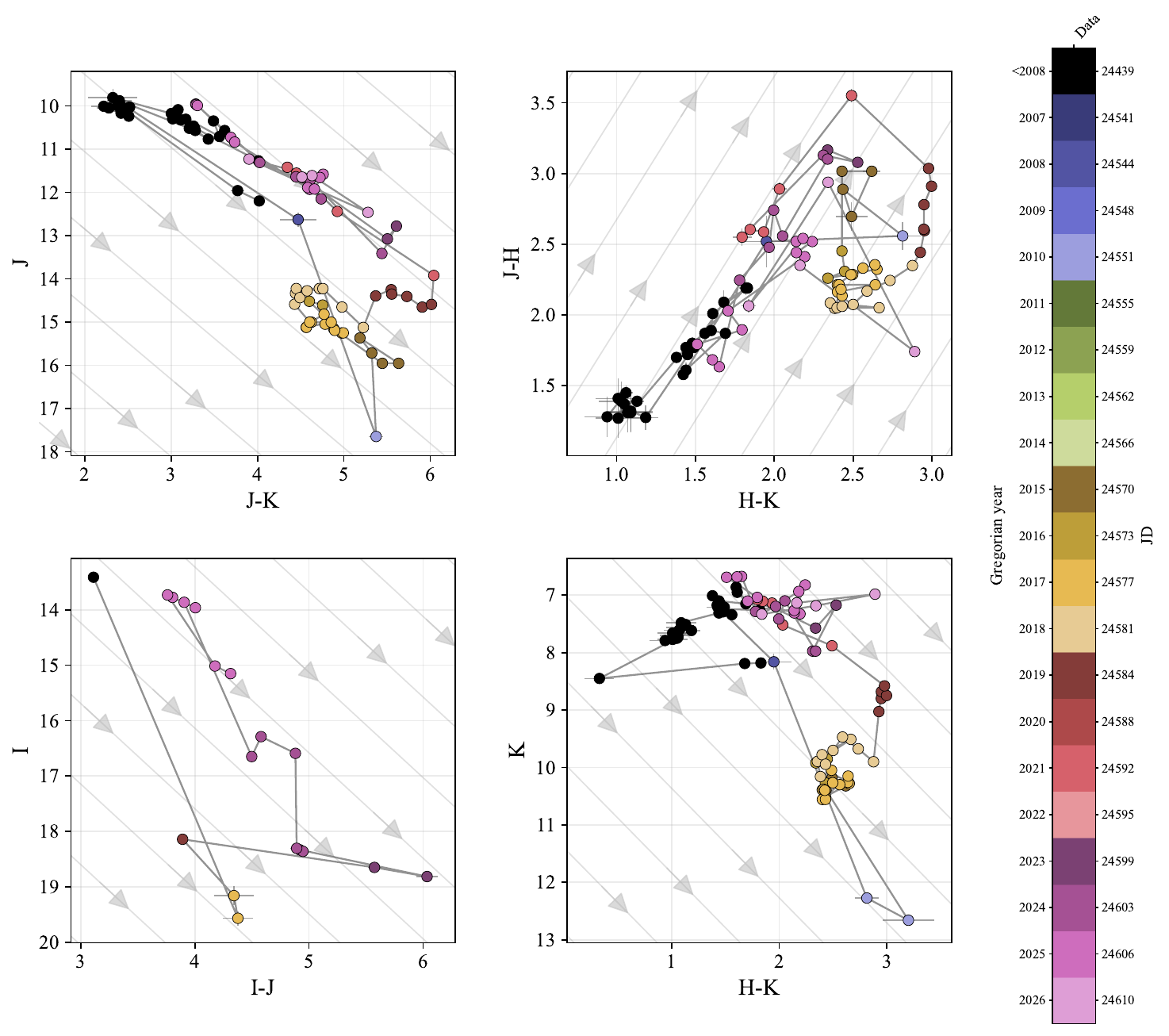}
\caption{Color-magnitude and color-color diagrams for V346\,Nor. Colored dots are measured photometry, while the gray lines with arrows mark the direction of reddening following \citet{cardelli1989}.}\label{fig:cmd}
\end{figure*}

In Fig.~\ref{fig:cmd}, we plotted color-magnitude and color-color diagrams, where data points from various observing years are displayed with different colored symbols. For comparison, we also plotted the direction of varying extinction using the reddening law of \citet{cardelli1989} with a total-to-selective absorption coefficient of $R=3.1$.

V346\,Nor follows a complex path in these diagrams. At certain times, it moves approximately along the reddening path, such as before 2008 (as already noted by \citealt{kospal2020}), during the sharp brightening in 2021, or in 2024-2025. At other times, it behaves very differently, such as during the deep 2010-2011 minimum, which was due to a temporary halt in the accretion rate \citep{kraus2016,kospal2017a}. In 2016-2017, it again seems to be very faint for its color, while in 2019 it moves in these graphs like never before due to the anti-correlating behavior in the $IJ$ and $HK_S$ bands. This complicated behavior suggests a combination of underlying physical causes for the variability, including varying line-of-sight extinction and varying accretion rate.

\subsection{Evolution of the infrared continuum slope}
\label{sec:sed}

Figure~\ref{fig:spec} shows our flux-calibrated optical--near-infrared--mid-infrared spectra of V346\,Nor covering the 2015-2026 time period. The photometric brightness and color changes are also reflected in the variations of the flux levels and shapes/slopes of the spectra. The XSHOOTER spectra from 2015 \citep{kospal2020} and the SpeX spectrum from 2015 \citep{connelley2018} caught the object in a faint state. At that time, V346\,Nor was already past the deep minimum due to the significant decrease in accretion rate in 2010. Nevertheless, in 2015, the object was still much fainter than it has been since then in the whole $1-4\,\mu$m wavelength range during our monitoring. In 2023 July, SpeX measured it about $2.9\,$mag brighter in all the $J$, $H$, and $K_S$ bands than in 2015. In 2024 February, it faded by $0.4\,$mag in all near-infrared bands. In 2024 September, the brightening was wavelength-dependent: $1.3\,$mag in $J$, $0.9\,$mag in $H$, but only $0.6\,$mag in $K_S$. In the following month, by 2024 October, 
it brightened again by another $0.8,0.3,0.1\,$mag in $JHK_S$, respectively.

The two SpeX spectra from 2025 March and April caught V346\,Nor in the maximum at $J=10.0\,$mag, $H=8.2\,$mag, and $K=6.7\,$mag. With this, it became brighter than ever recorded before, about $0.2\,$mag brighter in all near-infrared bands than at the previous peak in 1992 May \citep{prusti1993}. The object became bluer during this brightening: magnitude changes between 2024 September and 2025 April are $\Delta J=2.1\,$mag, $\Delta H=1.3\,$mag, $\Delta K=0.8\,$mag. Changes were much less significant at longer wavelengths: only $0.3\,$mag in the $3-4\,\mu$m range ($L$ band) between 2023 July and 2025 April based on the SpeX LXD\_long grism measurements. The next two SpeX spectra from 2025 June-July were taken after the steep fading, during the current intermediate flux level plateau. This fading was again color-dependent: $\Delta J=1.9\,$mag, $\Delta H=1.2\,$mag, $\Delta K=0.7\,$mag, $\Delta L=0.3\,$mag between 2025 April and June. Our latest SpeX measurement, from 2026 July, observed the object in a similar state to the previous one from 2025 July.

Two recent low-resolution, 0.75 -- 5$\,\mu$m spectra are available for V346\,Nor from SPHEREx, one from 2025 August - September, and one from 2026 February - April, both partly saturated. Focusing on the non-saturated parts of the SPHEREx spectra ($<2.4\,\mu$m), we can conclude that they are well consistent with the range of fluxes and spectral slopes displayed by our latest SpeX spectra from 2025-2026.

As the ratio between the observed near-infrared magnitude changes $\Delta{J}/\Delta{K_S}$ is close to the canonical value of extinction ratio for interstellar dust $A_J/A_K = 2.5$ \citep{rieke1985, cardelli1989, indebetouw2005}, we checked if the observed spectral shape variations can be explained by changing circumstellar extinction. Using the interstellar reddening law from \citet{cardelli1989}, we attempted to deredden each spectrum with varying values of $A_V$ until it best matched the 2025-Apr-15 spectrum. We found that the spectra from 2015 could not be dereddened to match those from 2025-Apr-15. To a lesser extent, but the same is true for the spectra from 2023-Jul-22 and 2024-Feb-24. For the other spectra, we found that a correction by $\Delta A_V = 7.0\,$mag in 2024-Sep-12, 4.7\,mag in 2024-Oct-10, 0.2\,mag in 2025-Mar-31, 6.4\,mag in 2025-Jun-21, 5.2\,mag in 2025-Jul-05, and 5.4\,mag in 2026-Jul-04 can reproduce the flux and shape of the maximal spectrum from 2025 April to within 10-20\%. The $\Delta A_V$ values listed here correspond to total-to-selective absorption coefficient $R=3.1$ (typical for ISM-like dust). The corresponding values for $R=5.5$ (typical for  dense molecular clouds or star-forming environments) are: $\Delta A_V = 5.9\,$mag in 2024-Sep-12, 4.0\,mag in 2024-Oct-10, 0.2\,mag in 2025-Mar-31, 5.4\,mag in 2025-Jun-21, 4.4\,mag in 2025-Jul-05, and 4.6\,mag in 2026-Jul-04. This suggests that, in accordance with the photometric analysis, V346\,Nor was undergoing extinction changes between 2024 September and 2025 September, and the decrease in extinction was responsible for the source's plateau of peak brightness seen in the OGLE $I_C$ and Atlas $o$ light curves in 2025 February -- May (Figs.\ref{fig:light} and \ref{fig:light_nir}). However, our near-infrared observations cannot constrain the value of the total-to-selective absorption coefficient, $R$, because the extinction curves for different $R$ values run parallel in the near-infrared. This is illustrated in Fig.~\ref{fig:avcorr} in the appendix.

\subsection{Evolution of spectral features}

\begin{figure*}
\centering
\includegraphics[width=\columnwidth,angle=0,page=1]{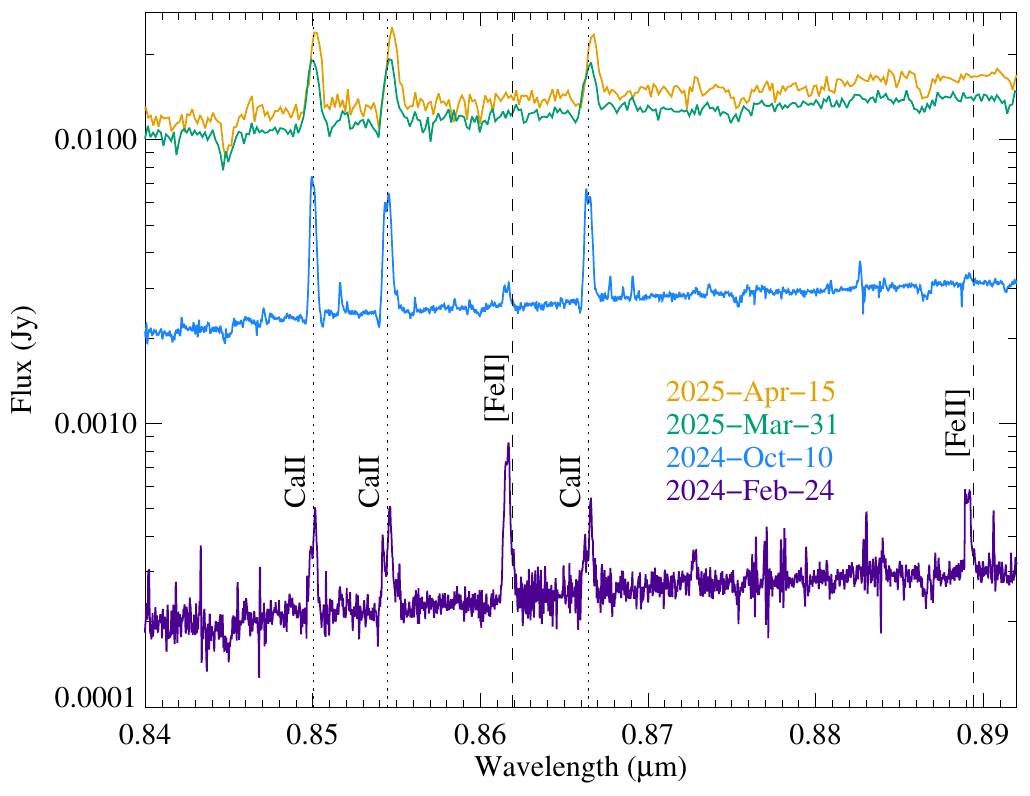}
\includegraphics[width=\columnwidth,angle=0,page=2]{test.pdf}
\includegraphics[width=\columnwidth,angle=0,page=1]{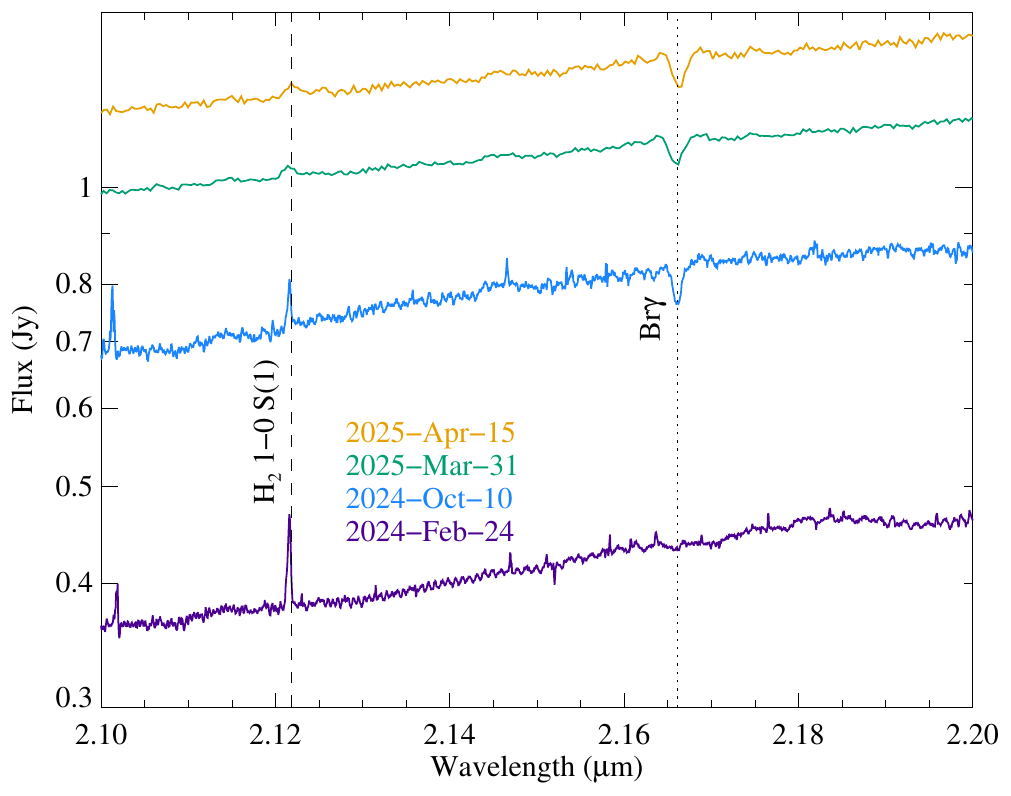}
\includegraphics[width=\columnwidth,angle=0,page=2]{test3.pdf}
\caption{Parts of selected spectra of V346\,Nor. Colors are the same as in Fig.~\ref{fig:spec}. The left panels display the original spectra at native resolution using a logarithmic y axis scale. The right panels show continuum-subtracted spectra using linear y axis scale. In the right panel, the XSHOOTER spectra (from 2024) were convolved with a Gaussian to match the spectral resolution of the SpeX spectra (from 2025). No extinction correction was applied to the spectra. Permitted/forbidden lines are marked with dotted/dashed lines.}\label{fig:lines1}
\end{figure*}

\begin{figure*}
\centering
\includegraphics[width=0.33\textwidth,angle=0,page=1]{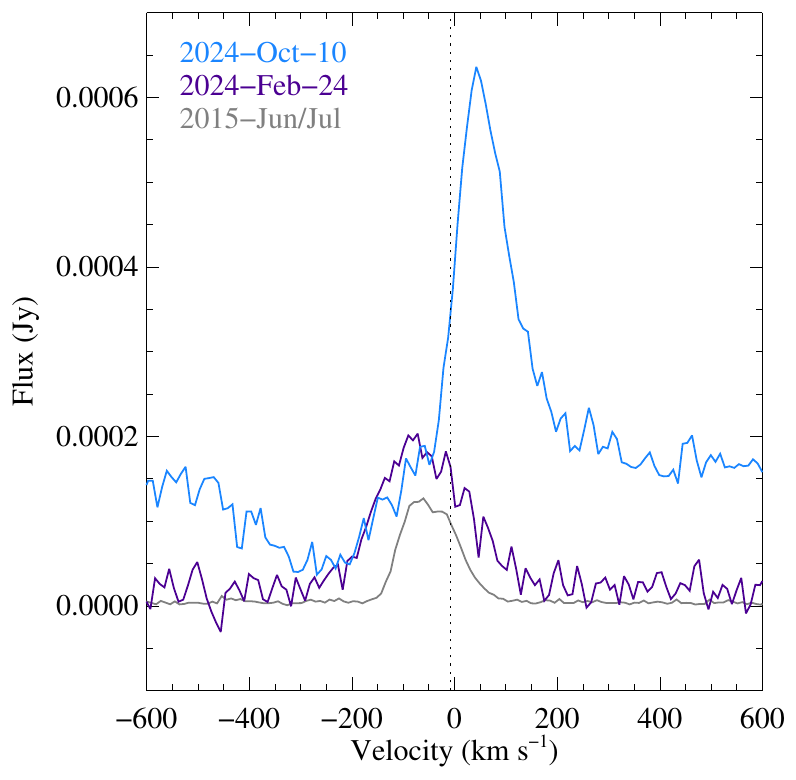}
\includegraphics[width=0.315\textwidth,angle=0,page=1]{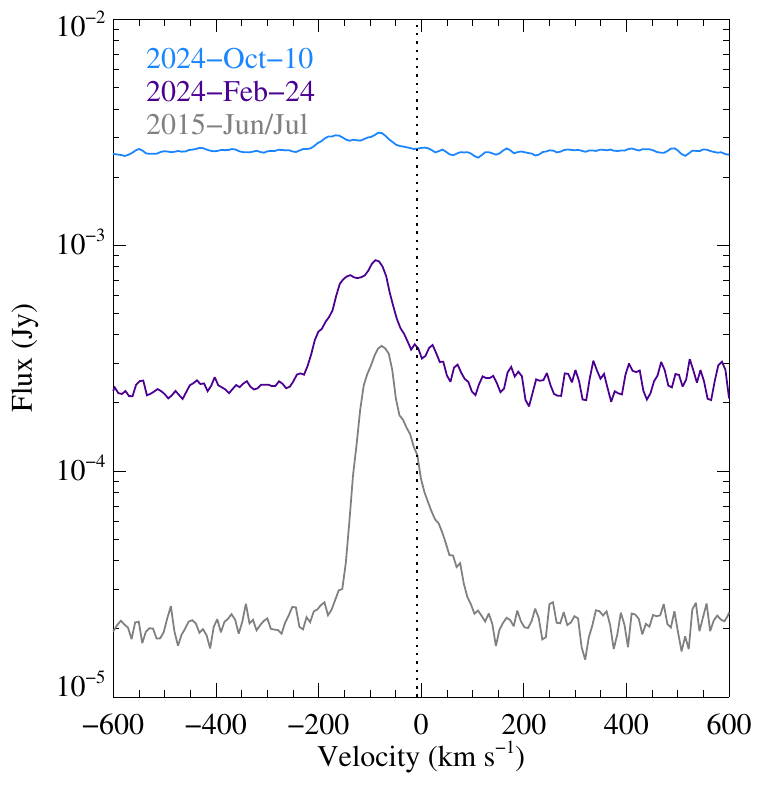}
\includegraphics[width=0.32\textwidth,angle=0,page=1]{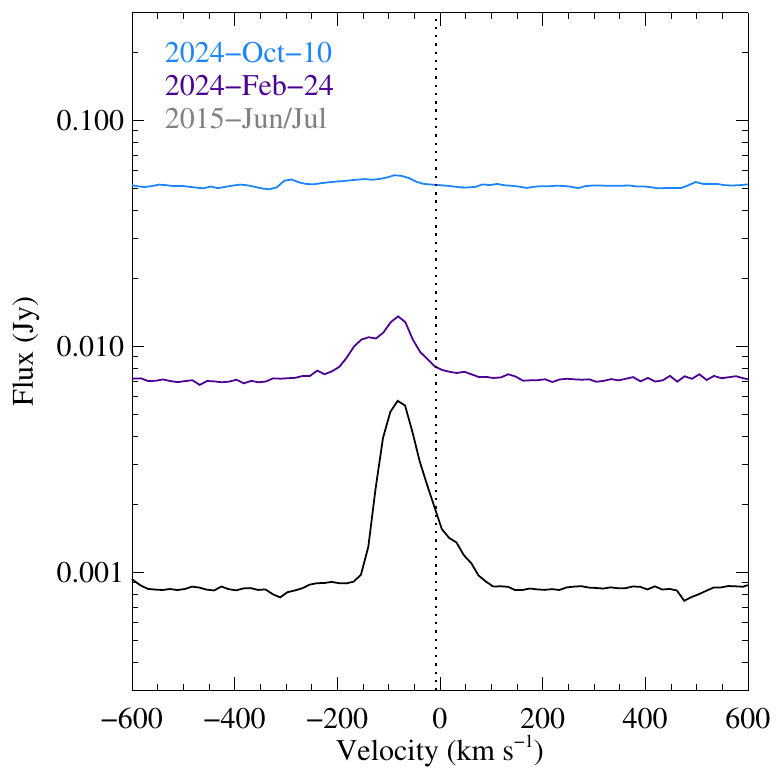}
\includegraphics[width=0.32\textwidth,angle=0,page=1]{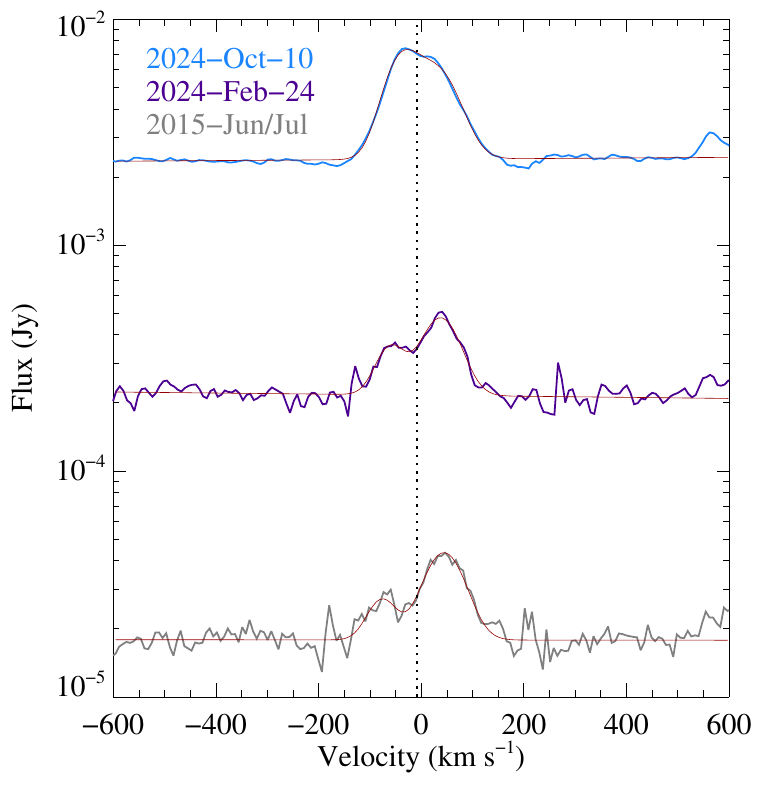}
\includegraphics[width=0.32\textwidth,angle=0,page=1]{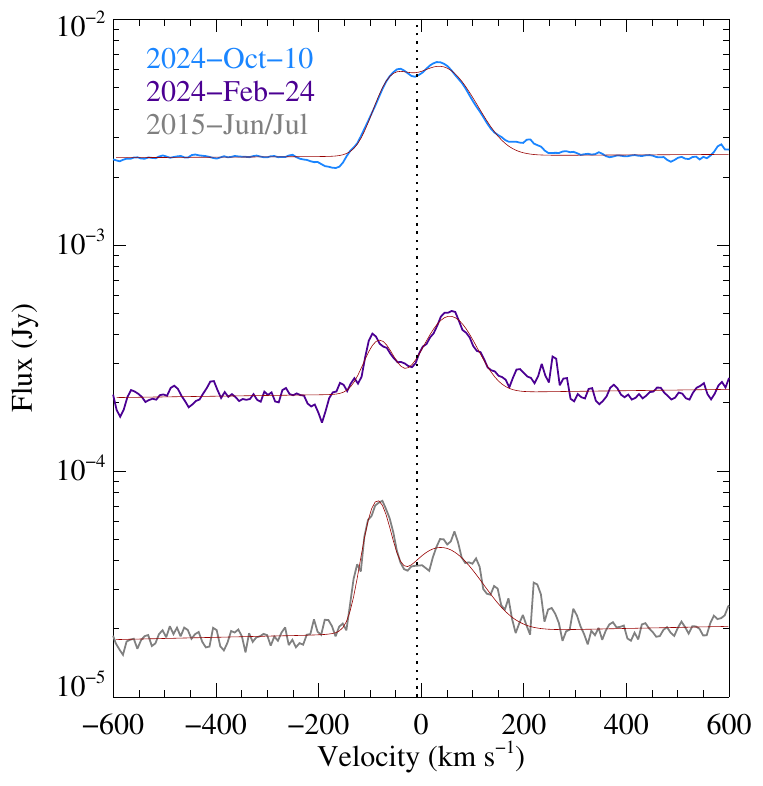}
\includegraphics[width=0.32\textwidth,angle=0,page=1]{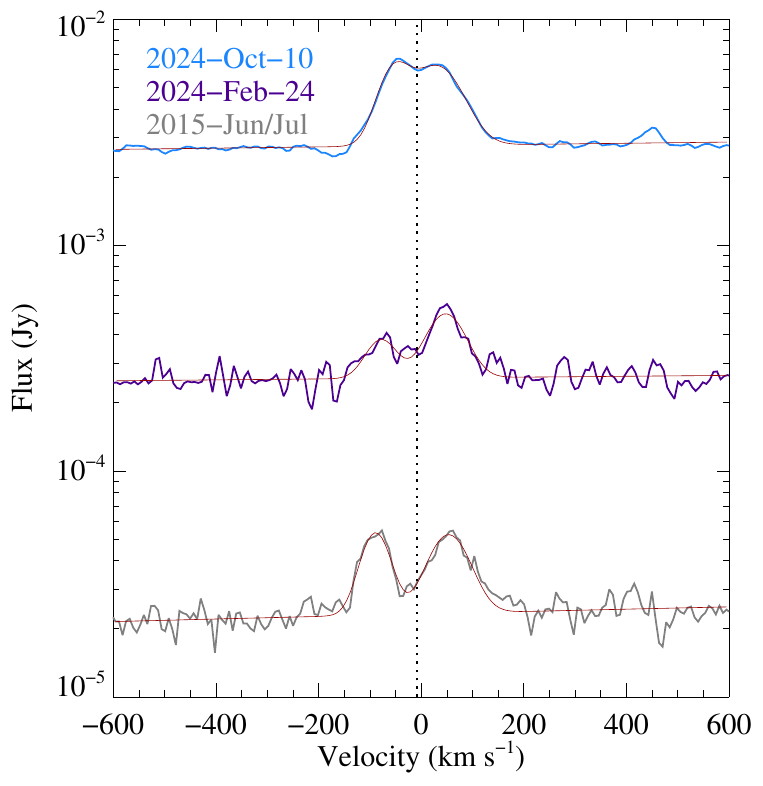}
\caption{Profiles of the H$\alpha$ (upper left), [FeII] 861.7\,nm (upper middle), [FeII] 1256\,nm (upper right), and the three CaII IRT lines (849.8\,nm to the left, 854.2\,nm at the center, 866.2\,nm to the right). Vertical dotted line marks the systemic velocity of $-7.8$\,km/s of V346\,Nor. Double Gaussian fits to the CaII IRT lines are plotted with thin red curves.}\label{fig:lines2}
\end{figure*}

\citet{kospal2020} analyzed in detail the spectral features observed toward V346\,Nor in a deep, high signal-to-noise ratio VLT/XSHOOTER spectrum from 2015 June/July. They found numerous forbidden atomic lines of [OI], [NII], [SII], [FeII], [CaII], [NiII], [CrII], [CI], [NI], [VI], as well as several forbidden molecular H$_2$ rovibrational lines. These lines are still visible in our newer spectra if the signal-to-noise allows it. As examples, two [FeII] lines are displayed in the upper row of Fig.~\ref{fig:lines1} and the H$_2$ 1--0 S(1) line is displayed in the lower row of Fig.~\ref{fig:lines1}. As before, the [FeII] lines are blueshifted (see also Fig.~\ref{fig:lines2}), while H$_2$ is centered at the systemic velocity (cf.~Fig.~7 of \citealt{kospal2020}), and there is no large variation in the strength of these forbidden lines; they are constant within about 35\%.

For permitted transitions, the Ca infrared triplet (IRT) is well visible in emission in all spectra with high enough signal-to-noise ratio in this wavelength range, i.e. with SpeX in 2025 April and May, and with XSHOOTER in 2024 February and October. These are shown in the upper row of Fig.~\ref{fig:lines1}. They display significant variations, strengthening with each epoch. This seems to correlate with the brightness of the continuum and prompted us to apply extinction correction to the spectra. We re-fitted the $A_V$ values that transform the 2024 February, October, and 2025 March spectra to match the continuum level of the 2025 April spectrum, as in Sec.~\ref{sec:sed}, but optimized the fitting to the $0.8-0.9\,\mu$m wavelength range. Then we applied dereddening using $A_V = 8.83, 3.74$, and $0.04\,$mag to these three spectra, respectively. This correction left the Ca IRT lines in the 2025 March almost unchanged, increased the lines in the 2024 October spectrum to twice stronger than in 2025 April, and brought the lines in the 2024 February spectrum to an almost exact match to those in 2025 April. Therefore, we can conclude that there is a real, physical variability in the CaII IRT lines in V346\,Nor.

\citet{kospal2020} found that the CaII IRT lines have two clearly distinct velocity components (also seen in Fig.~\ref{fig:lines2}), one blueshifted and one redshifted compared to the systemic velocity of V346\,Nor. The line ratios showed that the blueshifted component is marginally optically thin and is probably associated with shocked material in the jet just like the forbidden emission lines, while the redshifted component is optically thick. There was no mention of the origin of this component in that paper, but the optical depth and the redshifted velocity indicates that the calcium gas is moving away from the observer and toward the stellar surface. This suggests that this component may be associated with an accretion funnel flow, which would be an unexpected, surprising feature in a FUor-type young eruptive star, or with localized shocks or dense clumps falling onto the star in a boundary layer, which would be more usual for a FUor \citep{kwan2011}.

Our two new XSHOOTER spectra confirm that the CaII IRT lines are still double-peaked in 2024 February and October, but the line ratios and peak velocities changed. We fitted the sum of two Gaussian components to each line in the three XSHOOTER spectra (Fig.~\ref{fig:lines2}, lower row). By averaging the velocities of the fitted Gaussians for the three lines, we found that the blueshifted component peaked at heliocentric velocity of $-84\pm4\,$km/s, $-73\pm7\,$km/s, and $-51\pm6\,$km/s, while the redshifted component peaked at $+45\pm5\,$km/s, $+47\pm5\,$km/s, and $+33\pm2\,$km/s in 2015 June/July, 2024 February, and 2024 October, respectively. Therefore, there is a hint for changing velocities in the jet and possible accretion flow. The line ratios were 1:3.31:5.50 and 1:1.04:1.13 for the blueshifted and redshifted components in 2015. This also changed, becoming 1:1.15:0.91 and 1:1.00:0.91 for the blueshifted and redshifted components in 2024 February and 1:0.57:0.77 and 1:1.06:0.98 in 2024 October, suggesting that in 2024, both components were optically thick.

Among the HI lines, Pa$\beta$ was not detected in the 2015 XSHOOTER spectrum by \citet{kospal2020}, while Br$\gamma$ was not mentioned in that paper. By re-checking that spectrum, there is a hint for a weak Br$\gamma$ absorption. The SpeX spectrum from 2015 July \citep{connelley2018} does not show this, probably due to lower signal-to-noise ratio. V346\,Nor displays a smooth continuum around the Br$\gamma$ wavelength in 2023 July (SpeX), 2024 February (XSHOOTER), and 2024 September (SpeX). From 2024 October until the most recent spectrum in 2026 July, it consistently shows Br$\gamma$ absorption, with varying strength. Some of these can be seen in the lower row of Fig.~\ref{fig:lines1}. The appearance of the Br$\gamma$ line over our monitoring is summarized in Tab.~\ref{tab:spec}

Our SpeX spectra do not cover the H$\alpha$ line, so this can be studied only with XSHOOTER. The upper left panel of Fig.~\ref{fig:lines2} shows the H$\alpha$ line profiles. In 2015, the line had a blueshifted component and a zero velocity component, interpreted as combination of a high-velocity collimated jet component and a low-velocity wind component by \citet{kospal2020}. In 2024 February, the signal-to-noise ratio does not allow us the distinguish between these two components, but the profile is clearly dominated by blueshifted emission extending out to at least $-200\,$km/s. By 2024 October, the profile of the H$\alpha$ line became strikingly different, displaying a characteristic P\,Cygni profile with strong, slightly redshifted emission peak and a broad, blueshifted absorption trough extending out to about $-450\,$km/s, suggesting the presence of strong outflowing wind. This profile is reminiscent of the H$\alpha$ line observed in the classical FUor V1057\,Cyg (cf.~Fig.~9 in \citealt{szabo2021}).

Not only the CaII IRT and the H$\alpha$, but also the molecular CO overtone bandhead feature is highly variable in V346\,Nor. Sometimes it appears in emission, like in 1993 April/May, when \citet{reipurth1997} observed V346\,Nor; sometimes it is in absorption, like in 2015 June-July \citep{kospal2020}. \citet{reipurth1997} theorized that the weak CO bandhead emission may be due to a neutral disk wind. We observe weak CO emission in 2025-Mar-31, a continuum spectrum in 2023-Jul-22 and 2025-Apr-15, and absorption at all other epochs. This is summarized in Tab.~\ref{tab:spec}. Whether the feature is in emission, absorption, or non-detected, does not seem to correlate with other spectral features (such as the Br$\gamma$) or with the continuum brightness. 

\citet{graham1998} detected the $3.1\,\mu$m water ice absorption feature in V346\,Nor in 1995 June. The feature had a maximal optical depth of $\tau = 0.46$ and from its shape, the author concluded that the ice absorption does not originate in the dense circumstellar envelope of V346\,Nor, but rather comes from quiescent intracloud material along the line-of-sight to the star. We measured the optical depth of the $3.1\,\mu$m water ice feature in the 2015 July and 2025 April IRTF/SpeX spectra (see also the upper panel of Fig.~\ref{fig:seds}), and found $\tau = 0.15$ and $\tau = 0.26$, respectively. However, these numbers are highly uncertain due to the low signal-to-noise of the spectra in this wavelength range, and uncertainties associated with defining the underlying continuum due to the total atmospheric opacity window in the $2.5 - 2.9\,\mu$m wavelength range. Therefore, while there is a hint for the variability of the $3.1\,\mu$m water ice absorption feature, follow-up observations would be needed to confirm this tentative result.

\begin{table*}
\caption{Variability of spectral features in V346\,Nor}
\label{tab:spec}
\centering
\begin{tabular}{c c c c c c c c c}
\hline\hline
Date         & Telescope / & H$\alpha$ & Ca IRT   & H$_2$ 2.12$\,\mu$m & Br $\gamma$ & CO bandhead & $M_{\star}\dot{M}$ & $A_V$ \\
 & Instrument & & & & & & ($M_{\odot}^2\,$yr$^{-1}$) & (mag) \\
\hline
2015-Jun/Jul & X & emission & emission & emission & absorption & absorption & $8.5\times10^{-7}$ & 16.9 \\
2015-Jul-22  & I & ?        & ?        & emission & continuum  & absorption & $4.7\times10^{-7}$ & 14.0 \\
2023-Jul-22  & I & ?        & ?        & emission & continuum  & continuum  & $8.3\times10^{-5}$ & 25.2 \\
2024-Feb-24  & X & emission & emission & emission & continuum  & absorption & $3.8\times10^{-5}$ & 23.2 \\
2024-Sep-12  & I & ?        & ?        & emission & continuum  & absorption & $5.5\times10^{-5}$ & 20.4 \\
2024-Oct-10  & X & P Cygni  & emission & emission & absorption & absorption & $2.8\times10^{-5}$ & 15.0 \\
2025-Mar-31  & I & ?        & emission & emission & absorption & emission   & $3.9\times10^{-5}$ & 12.1 \\
2025-Apr-15  & I & ?        & emission & emission & absorption & continuum  & $4.0\times10^{-5}$ & 11.8 \\
2025-Jun-21  & I & ?        & ?        & emission & absorption & absorption & $4.0\times10^{-5}$ & 18.7 \\
2025-Jul-05  & I & ?        & ?        & emission & absorption & absorption & $5.0\times10^{-5}$ & 18.1 \\
2026-Jul-04  & I & ?        & ?        & emission & absorption & absorption & $2.9\times10^{-5}$ & 16.4 \\
\hline
\end{tabular}
  \tablefoot{`X' denotes VLT/XSHOOTER, `I' denotes IRTF/SpeX. `?' means that the spectral feature in question was not covered by the instrument or that the signal-to-noise ratio in that wavelength range was insufficient for a conclusive statement about detection. $M_{\star}\dot{M}$ and $A_V$ values are from our accretion disk model fitting, as explained in Sec.~\ref{sec:acc}.}
\end{table*}

\subsection{Accretion disk modeling}
\label{sec:acc}

To better understand the mechanisms underlying the eruptive behavior of V346\,Nor, we fitted our multi-epoch data with a steady optically thick and geometrically thin viscous accretion disk model, proposed for young eruptive stars by \citet{hk96}, and first used for V346\,Nor by \citet{kospal2017a}. Assuming a radially constant mass accretion rate, the radial temperature profile of the disk is $T^4_d(R) = \frac{3GM_{\star}\dot{M}}{8\pi{}R^3\sigma}\left[1-\left(\frac{R_{\star}}{R}\right)^{1/2}\right]$. The formal temperature reaches its maximum at $R=(49/36)R_{\star}\simeq1.36R_{\star}$; at smaller radii we set $T_{\rm d}(R) = T_{\rm max}$, to avoid non-physical decline towards the zero-torque inner boundary. The disk SED was calculated by integrating the blackbody emission of concentric annuli between $R_{\rm in}$ and $R_{\rm out}$. The model fluxes were reddened using the extinction law of \citet{cardelli1989}, adopting $R_V=3.1$. Following \citet{kospal2017a}, we fixed 
$R_{\star}=R_{\rm in}=3\,R_{\odot}$, $R_{\rm out}=2$\,au, the disk inclination to $i=40^{\circ}$, and the distance to 700\,pc. The exact value of $R_{\rm out}$ has negligible influence on the near-infrared fluxes. With these assumptions, the two free parameters at each epoch are $M_{\star}\dot{M}$ and $A_V$. More precisely, the disk temperature and luminosity constrain the product $M_{\star}\dot{M}$; the conversion to $\dot M$ therefore scales inversely with the adopted stellar mass.

We experimented with fitting the photometric SEDs over several wavelength ranges. At some epochs the complete optical-to-infrared SED could be reproduced by the accretion disk model, whereas at others clear optical and/or mid-infrared excess emission was present. We therefore adopted the $J$ and $H$ photometric measurements in our fitting procedure. This choice gives somewhat lower values of $M_{\star}\dot M$ and $A_V$ than those reported by \citet[][where the JHK range was fitted]{kospal2017a}, but reduces contamination by optical scattered light and by thermal emission from cooler dust at larger radii; in particular, the $K_S$ band often contains a substantial excess above the disk-only model. Each epoch was fitted by minimizing the chi-square in logarithmic flux density. A 5\% systematic uncertainty floor was added in quadrature to the measurement uncertainties. Because we used two data points for two fitted parameters, these photometric solutions are model-dependent inversions rather than over-constrained fits and do not by themselves provide a goodness-of-fit test. Fits including $K_S$ were therefore retained as a sensitivity check, rather than adopted as the fiducial solution.

We also fitted the available near-infrared spectra to test the photometric results and the degeneracy between extinction and the assumed inner radius. For the direct $JH$ comparison, the IRTF/SpeX SXD spectra were fitted over 1.20--1.60\,$\mu$m, with the 1.30--1.40\,$\mu$m telluric interval masked. The XSHOOTER spectra were fitted over 1.0--2.0\,$\mu$m after converting the supplied $F_{\lambda}$ values from erg\,s$^{-1}$\,cm$^{-2}$\,\AA$^{-1}$ to $F_{\nu}$ in Jy and masking the telluric bands. Because the XSHOOTER files contain no uncertainty spectra, we assigned a 10\% uncertainty before adding the common 5\% systematic floor. The spectral fits broadly confirmed the $JH$ photometric results, but did not remove the covariance between $R_{\rm in}$ and $A_V$. We therefore retained the fixed value $R_{\rm in}=3\,R_{\odot}$ throughout. Repeating the fits with a smaller ($2\,R_{\odot}$) or larger ($4\,R_{\odot}$) $R_{\rm in}$ resulted in systematically $4-7$\% lower or higher accretion rates and $0.5-1.1$\,mag higher or lower extinction, respectively. This degeneracy, together with possible deviations from a steady thin disk, should be kept in mind when interpreting the absolute parameter values; the relative temporal changes are more robust.

\begin{figure*}
\centering
\includegraphics[width=\textwidth]{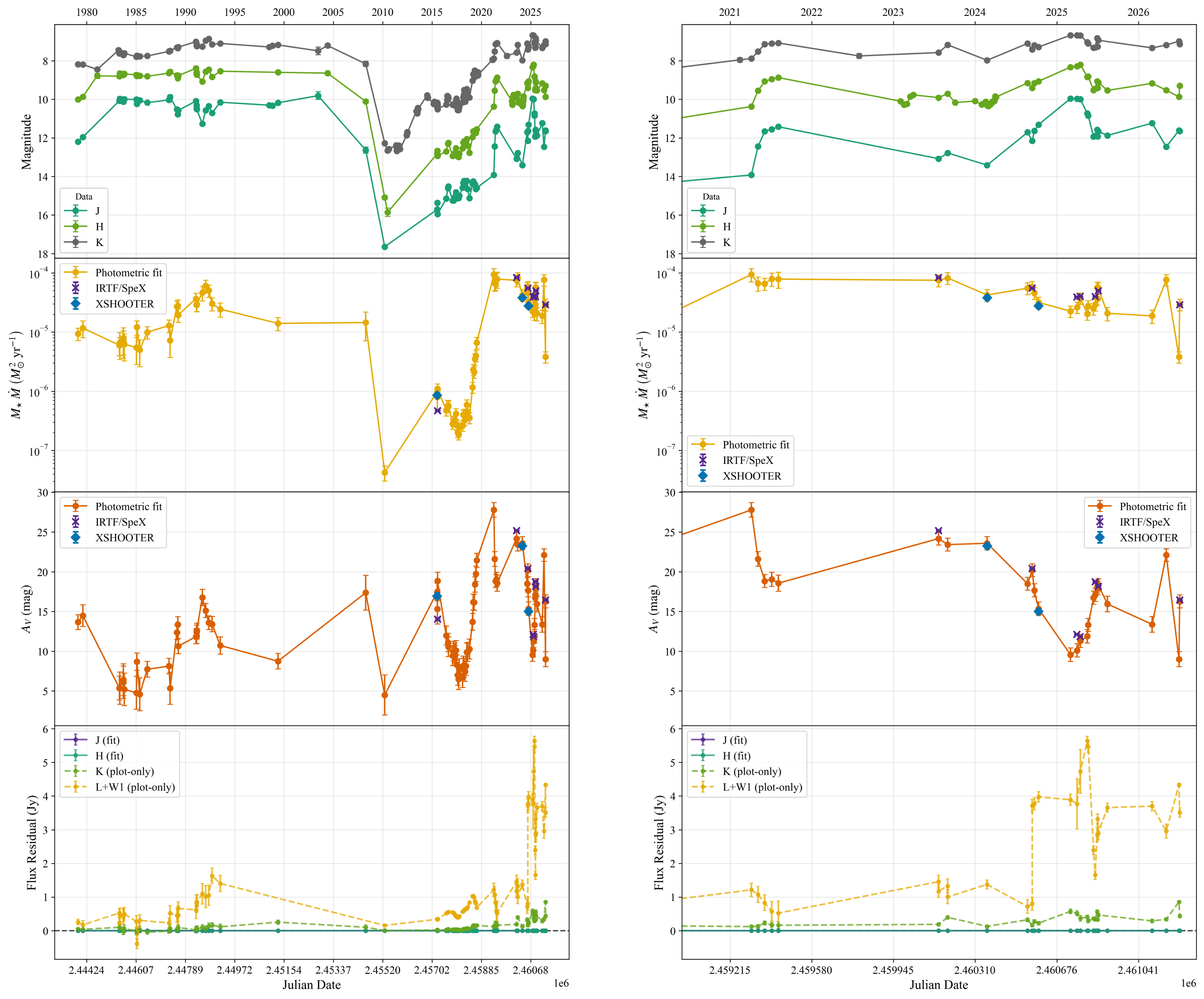}
\caption{Accretion disk modeling results for V346\,Nor. The left column shows the full 1979--2026 data set, while the right column zooms into the 2020--2026 time period. The upper row shows the observed $JHK_S$ photometry. The next two rows show the two fitted parameters: the product of the stellar mass and the accretion rate ($M_\star\dot{M}$), and the line-of-sight extinction ($A_V$). The last row shows the flux residuals in the $J$, $H$, $K_S$, $L'$, and $W1$ bands.}\label{fig:accdisk}
\end{figure*}

Figure~\ref{fig:accdisk} shows the resulting evolution of $M_{\star}\dot M$ and $A_V$. As discussed by \citet{kospal2020}, following its initial brightening in the late 1970s, V346\,Nor maintained $M_{\star}\dot M\sim10^{-5}\,M_{\odot}^{2}\,{\rm yr}^{-1}$, rising to approximately $6\times10^{-5}\,M_{\odot}^{2}\,{\rm yr}^{-1}$ during a local maximum in 1992 January. This was followed by a sharp decline to approximately $4\times10^{-8}\,M_{\odot}^{2}\,{\rm yr}^{-1}$ in 2010 March. By 2019 the source had gradually returned to its pre-fading state, while the fitted extinction also varied substantially. The accretion parameter subsequently increased to about $10^{-4}\,M_{\odot}^{2}\,{\rm yr}^{-1}$ in 2021.

From 2021 until 2025, $M_{\star}\dot M$ has been typically a few times $10^{-5}\,M_{\odot}^{2}\,{\rm yr}^{-1}$. The extinction declined from about 20\,mag in the second half of 2024 to a minimum of approximately 10\,mag in early 2025, while the accretion parameter changed much less. This temporary clearing can therefore account for a brightness maximum without requiring a comparable accretion burst.

The 2026 measurements do not show a simple monotonic continuation of the earlier rise. The fiducial $JH$ fits give $M_{\star}\dot M\simeq8\times10^{-5}\,M_{\odot}^{2}\,{\rm yr}^{-1}$ and $A_V\simeq22.1$\,mag on 2026 May 4, followed by substantially lower values in early July and a partial recovery within three days. The corresponding $JHK_S$ solutions are systematically higher because of the $K_S$-band excess. Thus, the most robust conclusion is that V346\,Nor reached exceptionally high accretion states during 2021--2026, but both accretion and line-of-sight extinction remained strongly variable.

Integrating the fiducial $JH$ accretion history by linear interpolation between all fitted epochs and fixing the stellar mass to $M_{\star} = 0.1\,M_{\odot}$ \citep{kospal2017c} gives a total accreted mass of approximately $1.8\times10^{-3}\,M_{\odot}$ over the observed 47-yr interval. This estimate is sensitive to sparse temporal sampling: excluding gaps longer than one year yields approximately $5.5\times10^{-4}\,M_{\odot}$ over the directly covered intervals, while the $JHK_S$ fits give a larger value of approximately $2.9\times10^{-3}\,M_{\odot}$. The integrated mass should therefore be quoted as model-dependent and bracketed by the sampling and wavelength-selection systematics. However, within the framework of the model considered here, the accreted mass is about 2\% of the stellar mass, which clarifies the order of magnitude of the accretion.

Our accretion disk fitting of the SEDs often exhibited excess radiation at mid-infrared wavelengths, which is probably associated with the emission of the dusty circumstellar medium surrounding the inner accretion disk. To investigate this excess, we subtracted the best-fitting accretion disk spectrum from the measured SED, de-reddened the residuals using the A$_V$ obtained for the disk, and fitted the residuals with a blackbody function. Most of the obtained excess temperature values are consistent with $T_{BB}=920\pm180$\,K. The most notable exception is $T_{BB}\sim550$\,K obtained on 2015 Jul 22. At this time the accretion rate was lower by at least one order of magnitude than at the peak luminosity (Fig.\,\ref{fig:accdisk}), thus the lower excess temperature may reflect a reduced heating of the circumstellar dust by the central source. The emitting area of the excess emission was rather constant over the whole outburst period with $R_{BB}=4.9\pm1.5$\,au, suggesting that the geometry of the circumstellar dust did not change considerably. At optical wavelengths we also measure an excess emission at most epochs, probably related to a scattered light component. The modeling of this emission is beyond the scope of this paper.

\section{Summary and conclusions}
\label{sec:sum}

In this paper, we presented a multi-wavelength monitoring campaign of the young eruptive star V346\,Nor, covering the period from 2023 to 2026. Our analysis, based on optical and near-infrared photometry, medium-resolution spectroscopy from VLT/XSHOOTER and IRTF/SpeX, and low-resolution spectra from the SPHEREx mission, allows us to characterize the nature of the 2024--2025 brightening event of V346\,Nor, which triggered the Gaia alert Gaia24chn.

Our photometric and spectrophotometric analysis reveals that the brightness and color variations of V346\,Nor are generally not the result of a single physical process, but rather a combination of varying accretion rates and changes in line-of-sight extinction. Our accretion disk modeling indicates that V346\,Nor has sustained a higher accretion rate since 2021 ($M_{\star}\dot{M} \sim 10^{-4} M_{\odot}^2$yr$^{-1}$) than during its previous historical peaks and it has accreted about 2\% of the stellar mass since its outburst in the late 1970s. The specific brightness maximum observed between 2025 February and May was primarily driven by a significant decrease in extinction, with $A_V$ dropping by approximately 10\,mag while the accretion rate remained relatively stable. This suggests that clearing of dust along our line of sight contributed to the observed increases in flux. The detection of significant infrared excess above the accretion disk model in the $K_S$, $L$, and $W1$ bands, which can be fitted at most epochs by a $920\pm180$\,K blackbody, points to the presence of a warm inner dust within a radius of $4.9\pm1.5$\,au.

The evolution of spectroscopic features provides further insight into the dynamic environment of the source. The appearance of a P\,Cygni profile in H$\alpha$ in 2024 October suggests the launch of a strong outflowing wind, reminiscent of classical FUors like V1057\,Cyg. Meanwhile, the variability of the CaII IRT lines and the transient appearance of CO bandhead emission indicate a highly unstable inner disk and accretion region. The presence of a redshifted component in the CaII IRT lines suggests that material is falling toward the stellar surface, possibly via dense clumps in a boundary layer. The persistence of forbidden emission lines, indicative of ongoing jet activity, suggests that V346\,Nor remains an active outflow source, even during periods of high accretion. 

In summary, V346\,Nor continues to exhibit the characteristics of an episodic accretion source. The 2024--2025 event demonstrates that outbursts in these objects can be a superposition of an actual increase in the mass accretion rate and the clearing of circumstellar dust. The current state of V346\,Nor, characterized by record-high accretion levels and strong wind activity, suggests that the object is currently in one of its most active phases since its original outburst in the 1970s. The Gaia-alerted brightening of V346\,Nor, and its subsequent complex evolution, adds to the growing body of evidence suggesting that FUor-like outbursts are not necessarily singular events but rather represent a continuous spectrum of accretion variability. 

\begin{acknowledgements}
We are grateful to the anonymous referee for constructive comments, which have helped to improve this manuscript. This project is based on observations collected at the European Southern Observatory under ESO programmes 095.C-0765, 0112.C-0440 (112.25PH), and 2113.C-5046 (113.28GX). The IRTF/SpeX data presented here were obtained at the Infrared Telescope Facility, which is operated by the University of Hawaii under contract 80HQTR24DA010 with the National Aeronautics and Space Administration. This publication makes use of data products from the Spectro-Photometer for the History of the Universe, Epoch of Reionization and Ices Explorer (SPHEREx), which is a joint project of the Jet Propulsion Laboratory and the California Institute of Technology, and is funded by the National Aeronautics and Space Administration. The OGLE project has received funding from the Polish National Science Centre grant OPUS-28 2024/55/B/ST9/00447 awarded to A.U. This work was also supported by ADVANCED 149943 and 2024-1.2.8-T\'ET-IPARI-CN-2025-00036 grants, which have been implemented with the support provided by the Ministry of Culture and Innovation of Hungary from the National Research, Development and Innovation Fund, financed under the NKKP ADVANCED and 2024-1.2.8-TÉT-IPARI-CN funding schemes. This work received funding from the Hungarian NKFIH grant K-147380. FCSM received financial support from the European Research Council (ERC) under the European Union's Horizon 2020 research and innovation programme (ERC Starting Grant ``Chemtrip'', grant agreement No 949278). EF has received funding from the European Research Council (ERC) via the ERC Synergy Grant ECOGAL (grant 855130). We acknowledge the Hungarian National Research, Development and Innovation Office grant OTKA FK 146023. We acknowledge support from the ESA PRODEX contract nr. 4000132054.
\end{acknowledgements}


\bibliographystyle{aa}
\bibliography{main}


\begin{appendix}
\nolinenumbers

\onecolumn

\section{Photometric data for V346\,Nor}

Table~\ref{tab:phot} shows the optical and near-infrared photometry for V346\,Nor presented and analyzed in this paper. Table~\ref{tab:wise} lists saturation-corrected WISE and NEOWISE photometry for V346\,Nor in seasonal (half-year) bins.

\begin{table*}[h!]
\caption{Optical and near-infrared photometry of V346\,Nor}
\label{tab:phot}
\centering
\tiny
\resizebox{\textwidth}{!}{
\begin{tabular}{c c c c c c c c c c c}
\hline\hline
Date & JD & Tel. & $g'$ & $r'$ & $i'$ & $z'$ & $J$ & $H$ & $K_s$ & $L'$\\
\hline
2015-Jun/Jul &            & VLT  & \dots          & \dots          & \dots          & \dots          & 15.693           & 12.840           & 10.385           & \dots \\
2015-Jul-22  & 2457225.75 & IRTF\tablefootmark{a} & \dots          & \dots          & \dots          & \dots            & 15.954$\pm$0.050 & 12.938$\pm$0.050 & 10.509$\pm$0.050 & \dots \\
2021-Feb-14  & 2459259.88 & VVVX & \dots          & \dots          & \dots          & \dots          & \dots            & \dots            &  7.950$\pm$0.118 & \dots \\
2021-Apr-06  & 2459310.60 & REM  & $>$22.20       & $>$21.78       & $>$21.06       & $>$20.78       & 13.922$\pm$0.041 & 10.371$\pm$0.023 &  7.881$\pm$0.016 & \dots \\
2021-May-06  & 2459340.65 & REM  & $>$22.39       & $>$21.98       & $>$22.05       & $>$21.81       & 12.442$\pm$0.015 &  9.551$\pm$0.043 &  7.518$\pm$0.013 & \dots \\
2021-Jun-05  & 2459370.70 & REM  & $>$21.21       & $>$20.92       & $>$20.30       & $>$19.93       & 11.657$\pm$0.011 &  9.070$\pm$0.015 &  7.137$\pm$0.015 & \dots \\
2021-Jul-06  & 2459401.56 & REM  & $>$22.54       & $>$21.91       & $>$21.51       & $>$21.27       & 11.556$\pm$0.021 &  8.952$\pm$0.033 &  7.105$\pm$0.011 & \dots \\
2021-Aug-05  & 2459431.75 & REM  & $>$22.42       & $>$21.87       & $>$21.50       & $>$21.26       & 11.421$\pm$0.012 &  8.871$\pm$0.059 &  7.075$\pm$0.012 & \dots \\
2021-Sep-04  & 2459461.56 & REM  & $>$22.54       & $>$21.95       & $>$21.81       & $>$21.58       & \dots            & \dots            &                  & \dots \\
2022-Aug-01  & 2459792.56 & VVVX & \dots          & \dots          & \dots          & \dots          & \dots            & \dots            &  7.748$\pm$0.113 & \dots \\
2023-Feb-02  & 2459977.81 & REM  & $>$20.47       & $>$20.82       & 18.86$\pm$0.13 & 16.92$\pm$0.28 & \dots            & 10.095$\pm$0.005 &  \dots           & \dots \\
2023-Feb-17  & 2459992.88 & REM  & \dots          & \dots          & \dots          & \dots          & \dots            & 10.254$\pm$0.012 &  \dots           & \dots \\
2023-Feb-18  & 2459993.88 & REM  & $>$19.94       & $>$19.82       & 18.88$\pm$0.13 & 17.71$\pm$0.29 & \dots            & 10.281$\pm$0.007 &  \dots           & \dots \\
2023-Feb-20  & 2459995.76 & REM  & $>$20.37       & $>$20.44       & 20.00$\pm$0.33 & 16.95$\pm$0.27 & \dots            & 10.253$\pm$0.007 &  \dots           & \dots \\
2023-Mar-03  & 2460006.86 & REM  & $>$19.68       & $>$20.72       & 19.34$\pm$0.45 & 17.90$\pm$0.51 & \dots            & 10.241$\pm$0.005 &  \dots           & \dots \\
2023-Mar-17  & 2460020.81 & REM  & $>$20.68       & $>$20.21       & 20.49$\pm$1.49 & 17.34$\pm$0.32 & \dots            &  9.856$\pm$0.005 &  \dots           & \dots \\
2023-Mar-31  & 2460034.83 & REM  & $>$20.82       & $>$20.69       & $>$20.26       & 19.31$\pm$2.56 & \dots            &  9.768$\pm$0.008 &  \dots           & \dots \\
2023-Jul-22  & 2460147.79 & IRTF & \dots          & \dots          & \dots          & \dots          & 13.077$\pm$0.015 & \dots            &  7.573$\pm$0.007 & \dots \\
2023-Sep-01  & 2460188.72 & IRTF & \dots          & \dots          & \dots          & \dots          & 12.781$\pm$0.010 &  9.702$\pm$0.007 &  7.173$\pm$0.008 & \dots \\
2023-Oct-04  & 2460221.56 & REM  & $>$21.79       & $>$21.40       & 20.73$\pm$0.26 & 17.43$\pm$0.11 & \dots            & 10.165$\pm$0.013 & \dots            & \dots \\
2024-Jan-01  & 2460310.84 & REM  & $>$20.56       & $>$20.34       & 18.83$\pm$0.14 & 18.68$\pm$1.32 & \dots            & 10.088$\pm$0.018 & \dots            & \dots \\
2024-Jan-31  & 2460340.81 & REM  & $>$21.42       & $>$21.53       & 19.46$\pm$0.12 & 17.81$\pm$0.11 & \dots            & 10.274$\pm$0.011 & \dots            & \dots \\
2024-Feb-02  & 2460342.83 & REM  & $>$20.97       & 20.89$\pm$1.18 & 18.91$\pm$0.17 & 18.46$\pm$0.45 & \dots            & 10.246$\pm$0.015 & \dots            & \dots \\
2024-Feb-13  & 2460353.78 & REM  & 21.69$\pm$1.23 & 20.19$\pm$0.40 & 19.13$\pm$0.19 & 17.48$\pm$0.21 & \dots            & 10.217$\pm$0.016 & \dots            & \dots \\
2024-Feb-18  & 2460358.85 & REM  & 20.09$\pm$0.20 & 20.36$\pm$0.32 & 19.30$\pm$0.17 & 17.78$\pm$0.20 & \dots            & 10.121$\pm$0.014 & \dots            & \dots \\
2024-Feb-25  & 2460365.83 & REM  & $>$19.74       & $>$20.18       & 18.58$\pm$0.87 & 18.92$\pm$0.98 & \dots            & 10.312$\pm$0.018 & \dots            & \dots \\
2024-Mar-01  & 2460370.89 & REM  & $>$21.24       & $>$21.09       & 19.26$\pm$0.25 & 18.07$\pm$0.48 & \dots            & 10.355$\pm$0.009 & \dots            & \dots \\
2024-Mar-07  & 2460376.89 & REM  & 22.19$\pm$1.29 & 20.02$\pm$0.26 & 19.21$\pm$0.19 & 18.63$\pm$0.56 & \dots            & 10.290$\pm$0.014 & \dots            & \dots \\
2024-Mar-12  & 2460381.89 & REM  & $>$21.52       & $>$21.54       & 19.00$\pm$0.16 & 17.50$\pm$0.16 & \dots            & 10.250$\pm$0.015 & \dots            & \dots \\
2024-Mar-17  & 2460386.89 & REM  & $>$21.97       & $>$21.42       & 19.47$\pm$0.31 & 17.65$\pm$0.18 & \dots            & 10.117$\pm$0.015 & \dots            & \dots \\
2024-Mar-22  & 2460391.89 & REM  & $>$21.89       & 21.22$\pm$1.13 & 19.05$\pm$0.16 & 18.07$\pm$0.30 & \dots            & 10.126$\pm$0.012 & \dots            & \dots \\
2024-Mar-27  & 2460396.89 & REM  & $>$20.23       & $>$20.27       & 18.35$\pm$0.32 & 17.39$\pm$0.40 & \dots            & 10.083$\pm$0.014 & \dots            & \dots \\
2024-Mar-28  & 2460397.85 & REM  & $>$19.60       & $>$19.49       & 18.07$\pm$0.41 & 17.47$\pm$0.69 & \dots            & 10.043$\pm$0.013 & \dots            & \dots \\
2024-Mar-31  & 2460400.86 & REM  & $>$21.00       & $>$20.57       & 18.92$\pm$0.13 & 17.62$\pm$0.12 & \dots            &  9.843$\pm$0.017 & \dots            & \dots \\
2024-Apr-04  & 2460404.85 & REM  & $>$22.09       & $>$21.70       & 19.18$\pm$0.12 & 17.27$\pm$0.02 & \dots            &  9.877$\pm$0.008 & \dots            & \dots \\
2024-Aug-23  & 2460545.61 & REM  & $>$21.27       & 19.87$\pm$0.27 & 17.46$\pm$0.01 & 15.95$\pm$0.02 & 11.710$\pm$0.014 &  9.152$\pm$0.010 &  7.099$\pm$0.016 & \dots \\
2024-Sep-12  & 2460565.74 & IRTF\tablefootmark{a} & \dots          & \dots          & \dots          & \dots            & 12.152$\pm$0.050 &  9.41$1\pm$0.050 &  7.415$\pm$0.050 & \dots \\
2024-Sep-22  & 2460575.60 & REM  & $>$20.53       & 19.79$\pm$0.30 & 17.62$\pm$0.06 & 15.79$\pm$0.06 & 11.640$\pm$0.008 &  9.162$\pm$0.007 &  7.195$\pm$0.009 & \dots \\
2025-Mar-02  & 2460737.16 & IRTF & \dots          & \dots          & \dots          & \dots          &  9.959$\pm$0.005 &  8.325$\pm$0.004 &  6.674$\pm$0.006 & \dots \\
2025-Mar-31  & 2460766.10 & IRTF & \dots          & \dots          & \dots          & \dots          &  9.972$\pm$0.004 &  8.289$\pm$0.004 &  6.681$\pm$0.005 & 4.412$\pm$0.148 \\
2025-Apr-15  & 2460781.10 & IRTF & \dots          & \dots          & \dots          & \dots          &  9.992$\pm$0.009 &  8.199$\pm$0.005 &  6.688$\pm$0.005 & 4.176$\pm$0.100 \\
2025-May-17  & 2460812.98 & IRTF & \dots          & \dots          & \dots          & \dots          & 10.729$\pm$0.006 &  8.834$\pm$0.005 &  7.038$\pm$0.005 & \dots \\
2025-May-21  & 2460816.98 & IRTF & \dots          & \dots          & \dots          & \dots          & 10.839$\pm$0.003 &  8.810$\pm$0.004 &  7.103$\pm$0.004 & 4.140$\pm$0.010 \\
2025-Jun-14  & 2460840.90 & IRTF & \dots          & \dots          & \dots          & \dots          & 11.931$\pm$0.009 &  9.519$\pm$0.007 &  7.326$\pm$0.006 & \dots \\
2025-Jun-21  & 2460847.83 & IRTF & \dots          & \dots          & \dots          & \dots          & 11.890$\pm$0.008 &  9.449$\pm$0.006 &  7.309$\pm$0.006 & \dots \\
2025-Jun-30  & 2460856.80 & IRTF & \dots          & \dots          & \dots          & \dots          & 11.582$\pm$0.004 &  9.063$\pm$0.004 &  6.822$\pm$0.004 & 4.565$\pm$0.013 \\
2025-Jul-02  & 2460858.84 & IRTF & \dots          & \dots          & \dots          & \dots          & 11.924$\pm$0.009 &  9.404$\pm$0.005 &  7.265$\pm$0.004 & \dots \\
2025-Jul-05  & 2460861.86 & IRTF & \dots          & \dots          & \dots          & \dots          & 11.658$\pm$0.003 &  9.117$\pm$0.003 &  6.935$\pm$0.003 & \dots \\
2026-May-04  & 2461165.05 & IRTF & \dots          & \dots          & \dots          & \dots          & 12.463$\pm$0.004 &  9.525$\pm$0.004 &  7.184$\pm$0.003 & 4.542$\pm$0.009 \\
2026-Jul-01  & 2461222.92 & IRTF & \dots          & \dots          & \dots          & \dots          & 11.613$\pm$0.003 &  9.872$\pm$0.003 &  6.982$\pm$0.003 & \dots \\
2026-Jul-04  & 2461225.81 & IRTF & \dots          & \dots          & \dots          & \dots          & 11.647$\pm$0.003 &  9.296$\pm$0.003 &  7.133$\pm$0.003 & \dots \\
\hline
\end{tabular}}
\tablefoot{
\tablefoottext{a}{Synthetic photometry from IRTF/SpeX spectra rather than photometry on the guidedog camera images.}
}
\end{table*}

\begin{table}[h!]
\caption{WISE and NEOWISE photometry of V346\,Nor}
\label{tab:wise}
\centering
\begin{tabular}{c c c}
\hline\hline
JD & $W1$ & $W2$ \\
\hline
2455258.86 & $8.203\pm0.026$ & $5.320\pm0.024$ \\
2455440.67 & $7.888\pm0.030$ & $5.224\pm0.046$ \\
2456723.47 & $7.100\pm0.057$ & $4.822\pm0.063$ \\
2456904.05 & $7.044\pm0.034$ & $4.806\pm0.027$ \\
2457085.38 & $7.236\pm0.019$ & $4.930\pm0.039$ \\
2457264.29 & $7.105\pm0.040$ & $4.654\pm0.042$ \\
2457452.10 & $6.981\pm0.067$ & $4.697\pm0.030$ \\
2457623.19 & $6.798\pm0.115$ & $4.449\pm0.063$ \\
2457816.62 & $6.891\pm0.068$ & $4.591\pm0.039$ \\
2457983.65 & $7.154\pm0.037$ & $4.674\pm0.038$ \\
2458181.21 & $6.785\pm0.095$ & $4.456\pm0.043$ \\
2458344.64 & $6.745\pm0.126$ & $4.470\pm0.061$ \\
2458548.13 & $6.060\pm0.167$ & $4.202\pm0.132$ \\
2458711.03 & $6.345\pm0.255$ & $4.326\pm0.047$ \\
2458912.20 & $5.846\pm0.133$ & $4.421\pm0.076$ \\
2459075.53 & $5.101\pm0.107$ & $4.004\pm0.136$ \\
2459276.32 & $5.112\pm0.076$ & $3.864\pm0.064$ \\
2459440.00 & $4.974\pm0.102$ & $4.024\pm0.135$ \\
2459643.35 & $4.838\pm0.113$ & $3.673\pm0.040$ \\
2459806.62 & $5.377\pm0.217$ & $3.851\pm0.068$ \\
2460007.50 & $5.048\pm0.067$ & $3.845\pm0.067$ \\
2460170.81 & $4.789\pm0.084$ & $3.515\pm0.044$ \\
\hline
\end{tabular}
\end{table}

\newpage
\clearpage

\section{Reddening-corrected IRTF spectra}
\label{sec:avcorr}

\begin{figure*}[h!]
\centering
\includegraphics[width=0.48\textwidth,angle=0,page=1]{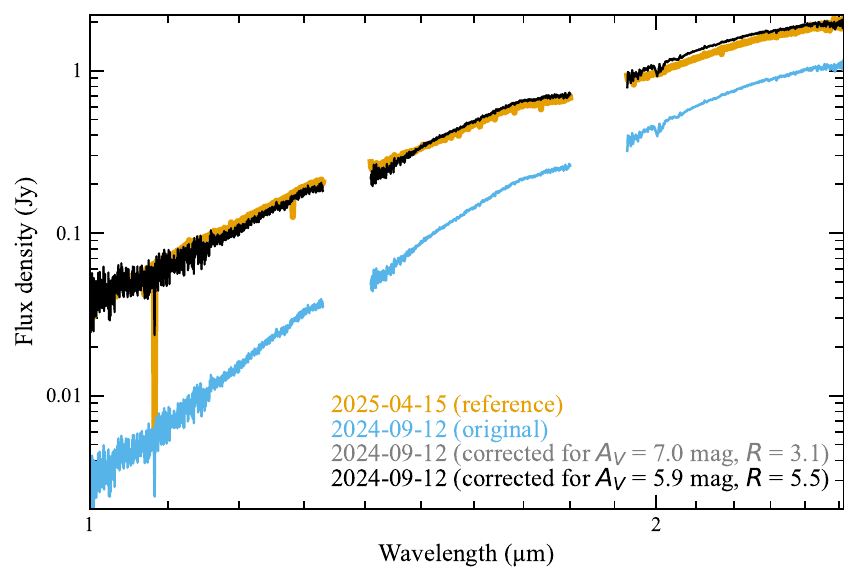}
\includegraphics[width=0.48\textwidth,angle=0,page=2]{avcorr.pdf}
\includegraphics[width=0.48\textwidth,angle=0,page=3]{avcorr.pdf}
\includegraphics[width=0.48\textwidth,angle=0,page=4]{avcorr.pdf}
\caption{Original and reddening-corrected IRTF/SpeX spectra of V346\,Nor at various observing epochs.}\label{fig:avcorr}
\end{figure*}

Figure~\ref{fig:avcorr} illustrates how spectra taken at certain epochs could be made to match with the one from 2025-Apr-15 by applying dereddening using the extinction law from \citet{cardelli1989}. Our data do not allow us to differentiate between the various total-to-selective absorption coefficients, because the extinction curves corresponding to $R=3.1$ and $R=5.5$ run parallel in the infrared, therefore, when considering the best-fitting solution, they result in the same infrared spectral slope but for different $A_V$ value. The gray (corresponding to $R=3.1$) and black (corresponding to $R=5.5$) curves in Fig.~\ref{fig:avcorr} differ by less than 0.3\% in the $1-2.5\,\mu$m wavelength range.

\newpage
\clearpage

\section{Checks on the SPHEREx absolute flux calibration and saturation}
\label{sec:spherex}

Since its launch in 2025 March and the start of its regular science operations in 2025 May, NASA's SPHEREx mission has been performing an all-sky spectral survey twice per year, covering wavelengths from 0.75$\,\mu$m to 5.0$\,\mu$m. SPHEREx has a 20\,cm effective aperture wide-field telescope, whose optical path is split by a dichroic beam splitter, which allows for simultaneous observations in short ($0.75-2.44\,\mu$m) and mid-wave ($2.40-5.01\,\mu$m) infrared bands. It uses a Linear Variable Filter (LVF) for spectroscopy, which provides a spectral resolution between 35 and 128. In the SPHEREx detector arrays, each pixel measures a different infrared color. The system combines data taken across several exposures to create full images and spectral data of any object in the sky.

According to the SPHEREx Explanatory Supplement\footnote{\href{https://irsa.ipac.caltech.edu/data/SPHEREx/docs/SPHEREx_Expsupp_QR.pdf}{https://irsa.ipac.caltech.edu/data/SPHEREx/docs/\\SPHEREx\_Expsupp\_QR.pdf}}, the absolute photometric calibration is based on $\sim2700$ measurements of 9 primary calibrator stars for the first Quick Release (QR1) and $\sim7400$ measurements of 11 primary calibrator stars for Quick Release 2 (QR2). These measurements were used to calculate a calibration function as a function of wavelength. These, together with the flat fields, are used to convert the images from e-/s to MJy/sr.

With the goal of checking the absolute flux calibration on more data and the saturation limits on actual satellite data, we used  106 calibrator stars from the CALSPEC catalog\footnote{\href{https://www.stsci.edu/hst/instrumentation/reference-data-for-calibration-and-tools/astronomical-catalogs/calspec}{https://www.stsci.edu/hst/instrumentation/reference-data-for-calibration-and-tools/astronomical-catalogs/calspec}} \citep{bohlin2020}. We downloaded their model spectra from the STScI website and downloaded SPHEREx measurements for them using the SPHEREx Spectrophotometry Tool available at IRSA\footnote{\href{https://irsa.ipac.caltech.edu/applications/spherex/tool-spectrophotometry}{https://irsa.ipac.caltech.edu/applications/spherex/tool-spectrophotometry}}. As the Spectrophotometry Tool does not account for proper motion, we downloaded data in two steps: once using the J2000 coordinates of the objects, then once we know when SPHEREx covered that location, we re-calculated their positions for those epochs using their proper motions and re-downloaded the data. 

\begin{figure*}[h!]
\centering
\includegraphics[width=\textwidth,angle=0,page=1]{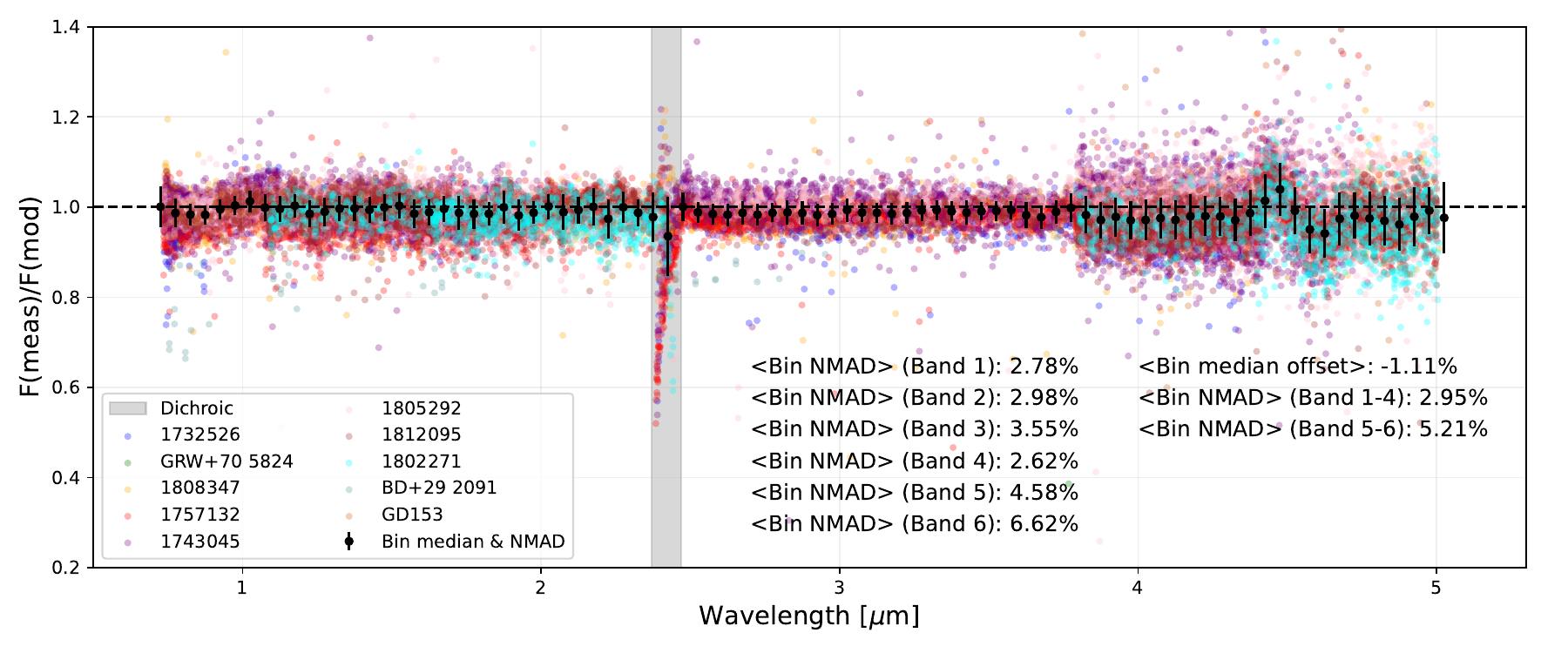}
\caption{Ratios of measured and modeled fluxes as a function of wavelength for 10 primary calibrator stars, based on Fig.~19 of the SPHEREx Explanatory Supplement. NMAD denotes the Normalized Median Absolute Deviation, a robust, outlyer-resistant statistical measure of the dispersion of data points. The median curve and the uncertainties are available as data behind the figure.}\label{fig:spherex_calib}
\end{figure*}

Figure~19 in the SPHEREx Explanatory Supplement shows the measured/predicted flux ratio for the 11 primary calibrators as a function of wavelength, suggesting no significant offsets (the median offset for all bands is 0.74\%) and calibration uncertainties of 2.8\% in Bands 1-4 and 5.3\% in Bands 5-6. In Fig.~\ref{fig:spherex_calib}, we reproduced this figure using all data available until 2025-Oct-02, 38\,000 individual measurements, for 10 out of the 11 primary calibrators (Feige 110 has no model spectrum in the CALSPEC database). In general, we found larger offsets but smaller uncertainties (Fig.~\ref{fig:spherex_calib}). We provide the median curve and the uncertainties in machine-readable format to allow the reader to apply these offset corrections.

\begin{figure*}[h!]
\centering
\includegraphics[width=\textwidth,angle=0,page=1]{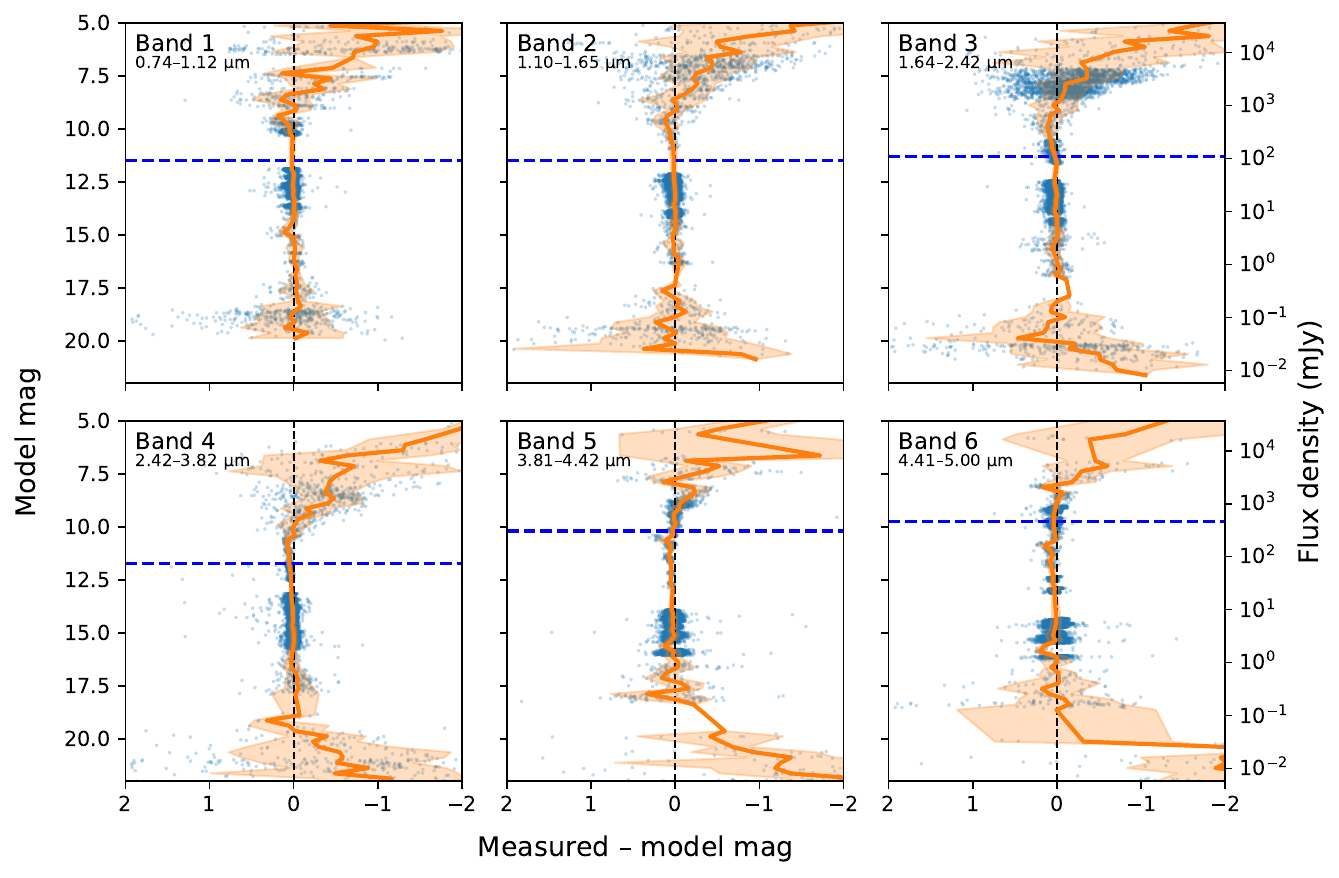}
\caption{Model magnitudes as a function of magnitude difference between the SPHEREx measurements and models for 83 of the studied 106 CALSPEC stars. Blue horizontal dashed lines mark the saturation limits based on a pre-launch performance model as given in the SPHEREx Explanatory Supplement. The thick orange line is the median curve, while the shaded orange region around it shows $\pm$ 1.4826 $\times$ MAD (median  absolute deviation), as a robust estimate of the how spread out the data are around its median.}\label{fig:spherex_saturation}
\end{figure*}

In terms of saturation, the SPHEREx Explanatory Supplement provides limits based on a pre-launch performance model. These saturation limits are 11.5\,mag in Bands 1-2 ($0.75-1.11\,\mu$m and $1.11-1.64\,\mu$m), 11.3\,mag in Band 3 ($1.64-2.42\,\mu$m), 11.71\,mag in Band 4 ($2.42-3.82\,\mu$m), 10.19\,mag in Band 5 ($3.82-4.42\,\mu$m), and 9.75\,mag in Band 6 ($4.42-5.00\,\mu$m). To check these numbers with actual data, we calculated the difference between the measured and model magnitudes for each star. After a visual inspection of these differences, we discarded 23 stars because they appeared to be outliers. For the remaining 83 stars, we plotted the model magnitudes as a function of the measured - model difference separately for each Band. This is shown in Fig.~\ref{fig:spherex_saturation}, again, using SPHEREx data available until 2025-Oct-02. In this figure, we also marked with blue horizontal dashed lines the saturation limits based on a pre-launch performance model.

Figure~\ref{fig:spherex_saturation} shows that below the saturation limit, but for not too faint stars, offsets and uncertainties are small as expected. At the faint limit, the uncertainties (i.e., the scatter or spread of data points) increase, and in Bands 3-6, there also seems to be an offset towards negative magnitudes, that is, the brightness of the faintest stars is overestimated. At the bright limit, offsets and scatter remain small for about 1\,mag above the official pre-launch saturation limits. Above this, the uncertainty increases and there is again an offset towards negative magnitudes in all bands: the brightness of saturated stars are also in general overestimated.

\newpage
\clearpage

\section{Accretion disk model fits to the SEDs of V346\,Nor}

\begin{figure*}[h!]
\centering
\includegraphics[width=\textwidth,angle=0,page=1]{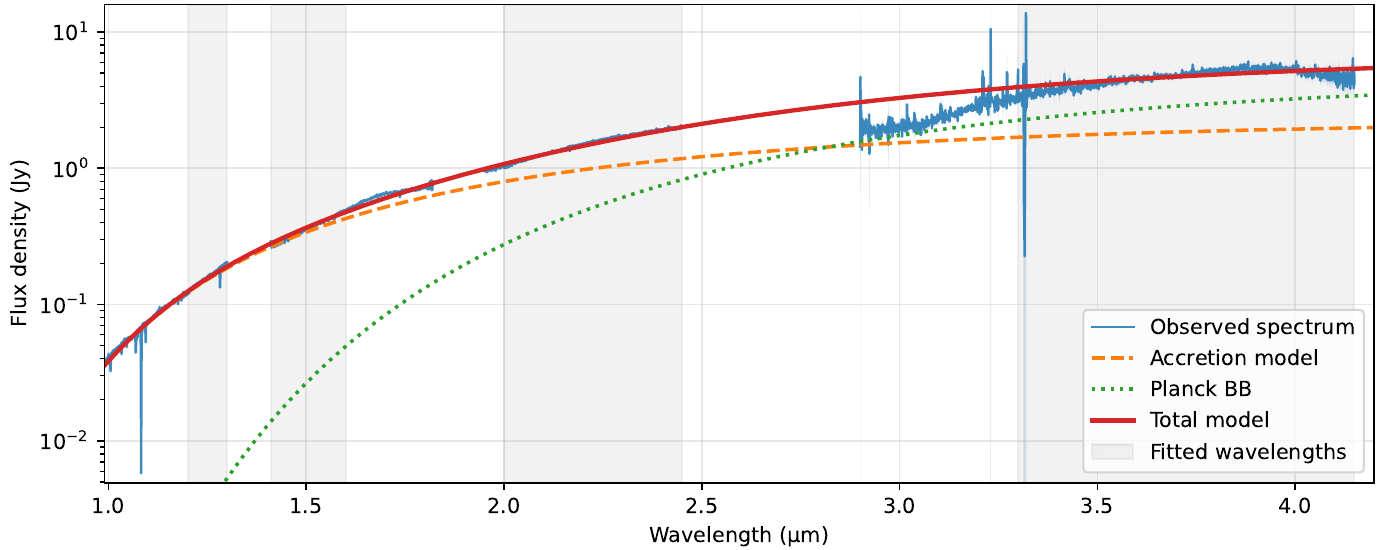}
\includegraphics[width=\textwidth,angle=0,page=1]{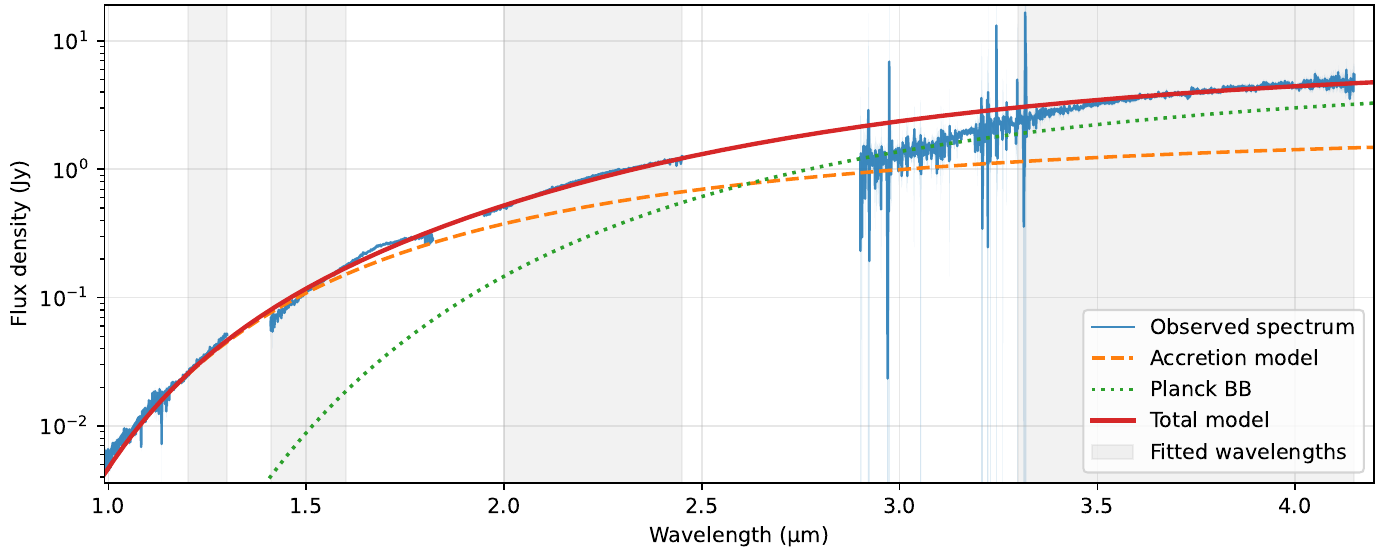}
\caption{Examples of accretion disk model fits to IRTF/SpeX spectra of V346\,Nor. Top: spectrum from 2025-Apr-15, bottom: spectrum from 2026-Jul-04. The blue curves are the observed spectra, yellow dashed curves are the accretion disk model, the green dotted curves are blackbody fits that account for the excess above the accretion disk. Solid red curves are the sum of the accretion disk and blackbody components. The gray shaded areas mark the fitting regions.}\label{fig:seds}
\end{figure*}

Figure~\ref{fig:seds} shows examples of our accretion disk model fits to IRTF/SpeX spectra. The significant deficit around $3\,\mu$m is due to water ice absorption. 

\end{appendix}

\end{document}